\documentclass[aps,
               prd,
               onecolumn,
               10pt,
               groupedaddress,
               superscriptaddress,
               draft,
               amsfonts,
               amssymb,
               amsmath,
               preprintnumbers,
               nobibnotes,
               nofootinbib,
               altaffilletter,
               showpacs,
               noshowkeys,
               floats,
               notitlepage,
               oneside,
               letterpaper,
               eqsecnum,
               citeautoscript,
               ]{revtex4}

\def\mca{{\mathcal A}}
\def\mcb{{\mathcal B}}
\def\mcl{{\mathcal L}}
\def\mcm{{\mathcal M}}
\def\mco{{\mathcal O}}
\def\mcr{{\mathcal R}}
\def\Lie{{\pounds}}

\begin{document}

%%%%%%%%%%%%%%%%%%%%%%%%%%%%%%%%%%%%%%%%%%%%%%%%%%%%%%%%%%%%%%%%%%%%%%%%
\title{
A doubly covariant formula of deficit angle and its application to six-dimensional braneworld
}

\author{
Yuuiti Sendouda
}\email[]{sendouda@utap.phys.s.u-tokyo.ac.jp}
\affiliation{
Department of Physics, Graduate School of Science, The University of Tokyo,
Hongo 7-3-1, Bunkyo-ku, Tokyo 113-0033, Japan
}
\author{
Shunichiro Kinoshita
}\email[]{kinoshita@utap.phys.s.u-tokyo.ac.jp}
\affiliation{
Department of Physics, Graduate School of Science,
The University of Tokyo,
Hongo 7-3-1, Bunkyo-ku, Tokyo 113-0033, Japan
}
\author{
Shinji Mukohyama
}\email[]{mukoyama@phys.s.u-tokyo.ac.jp}
\affiliation{
Department of Physics, Graduate School of Science,
The University of Tokyo,
Hongo 7-3-1, Bunkyo-ku, Tokyo 113-0033, Japan
}
\affiliation{
Research Center for the Early Universe,
Graduate School of Science, The University of Tokyo,
Hongo 7-3-1, Bunkyo-ku, Tokyo 113-0033, Japan
}

\date{
\today
}

\begin{abstract}
We reformulate boundary conditions for axisymmetric codimension-2 braneworlds in a way which is applicable to linear perturbation with various gauge conditions.
Our interest is in the thin brane limit and thus this scheme assumes that the perturbations are also axisymmetric and that the surface energy--momentum tensor of the brane is proportional to its induced metric.
An advantage of our scheme is that it allows much more freedom for convenient coordinate choices than the other methods.
This is because in our scheme, the coordinate system in the bulk and that on the brane are completely disentangled.
Therefore, the latter does not need to be a subset of the former and the brane does not need to stay at a fixed bulk coordinate position.
The boundary condition is manifestly doubly covariant:
it is invariant under gauge transformations in the bulk and at the same time covariant under those on the brane.
We take advantage of the double covariance when we analyze the linear perturbation of a particular model of six-dimensional braneworld with warped flux compactification.
\end{abstract}

%\keywords{}

\preprint{UTAP-565}
\preprint{RESCEU-21/06}

\maketitle

%%%%%%%%%%%%%%%%%%%%%%%%%%%%%%%%%%%%%%%%%%%%%%%%%%%%%%%%%%%%%%%%%%%%%%%%
\section{
Introduction
}
\label{sec:intro}

% Braneworld
The idea of braneworld \cite{Arkani-Hamed:1998rs,Arkani-Hamed:1998nn,Antoniadis:1998ig,Randall:1999ee,Randall:1999vf} has been attracting a great deal of physical interest as a scenario of our universe in higher-dimensional theories.
In this scenario our universe is thought to be a timelike surface, called brane, embedded in a higher-dimensional bulk spacetime.
The view of our universe as a `membrane' has been inspired by the string theory and each braneworld scenario is considered to reflect some of essential features of the fundamental theory.
In this sense, it is expected that various predictions from braneworld models should serve as tests to determine if the underlying theory is the true theory describing the universe or not.
In particular, possible deviation of gravity from the four-dimensional theory has been considered as one of the most promising probes of the higher-dimensional theory.

To compactify the extra dimensions is a key concept incorporated in higher-dimensional theories.
In general, properties of gravity deviate from the prediction of the four-dimensional Einstein theory at scales shorter than the compactification scale.
Experimental bound on the short distance deviation of gravity is not so strong and, in the context of braneworld scenarios, still allows the size of extra dimensions up to around $ 0.1~\mathrm{mm} $ long \cite{Hoyle:2000cv,Long:2002wn}.
To be more precise, one should consider this as a bound on the mass of moduli, i.e., four-dimensional scalar fields representing deformations of extra dimensions.
Even though the size of extra dimensions is smaller than $ 0.1~\mathrm{mm} $, light moduli would spoil the prediction of the four-dimensional Einstein theory at longer scales if the mass of the moduli were lighter than $ (0.1~\mathrm{mm})^{-1} $.
For this reason, there have been tremendous efforts to stabilize all moduli in string theory \cite{Kachru:2003aw} (for review, see \cite{Grana:2005jc} and references therein) and this has been one of the most important issues in string theory.

As the opposite extreme, another key in the study of higher-dimensional theories is the expansion dynamics of the four-dimensional universe, in which the unusual behavior of gravity can in principle be observed.
The non-standard expansion rate of the universe predicted by braneworld cosmologies serves as a test of the theories.
Furthermore, one of the most fundamental problems assigned to the fundamental theory is realization of cosmological inflation \cite{Guth:1980zm,Sato:1980yn}.
In this context, the use of the warped flux compactification has recently made it possible to investigate practical inflationary scenarios in the framework of the string theory \cite{Kachru:2003aw,Kachru:2003sx} (see \cite{Linde:2005dd} and references therein for review), and this can be considered as one of the most important achievements in very recent years.

For these reasons, the study of braneworlds has been getting more and more important.

% Double covariance
Now, the principle of general covariance is one of the most essential concepts in modern physics:
laws of physics must be independent of {\it a priori} geometry and the choice of coordinates, and, thus, covariant under general coordinate transformations.
In the context of braneworld scenarios, there are indeed two types of coordinate transformation.
One is on the brane and the other in the bulk.
Laws of physics in our world on the brane must be {\em doubly covariant}, i.e., covariant under both types of coordinate transformation.
To be more precise, they must be covariant under the coordinate transformation on the brane and invariant under the coordinate transformation in the bulk.

% Thickness
Physically speaking, a brane representing our universe is not just a mathematical surface but must be manifested as an extended object or a soliton in higher-dimensional theories.
Therefore it is in general necessary to specify properties (e.g., internal structure) of the extended object in order to describe dynamics of the braneworld.
Fortunately, in the codimension-1 case, it is possible to take the limit of zero thickness and obtain low-energy description of the dynamics without referring to microscopic properties.
The thin limit can be taken also in the codimension-2 case if the spacetime around a brane is axisymmetric and if the surface energy--momentum tensor includes vacuum energy, or tension, only.
With codimension more than $ 2 $, it is not possible to take the thin limit, keeping brane tension and gravitational coupling constant finite.
Indeed, for codimension more than $ 2 $, if we took the limit of zero thickness then gravity around the object would become so strong that it would form a black-hole horizon or a singularity.
Therefore, for the description of a brane with codimension more than $ 2 $, one must keep the brane thickness finite and specify the internal structure within the thickness in order to describe the dynamics of the brane.
From now on we shall restrict our attention to situations in which we can take the thin limit:
the general cases with codimension 1 and the special case with codimension 2.

% Codimension 1
In the codimension-1 case, Israel's junction condition \cite{Israel:1966rt} describes low-energy dynamics of braneworlds.
It is beautifully formulated in terms of geometrical quantities (induced metric and extrinsic curvature) and matter quantities (surface energy--momentum tensor) on the hypersurface.
However, a general scheme of perturbation of junction condition had not been formulated in a doubly covariant fashion until \cite{Mukohyama:2000ga}.
One of the obstacles was, to our knowledge, the fact that the coordinate position of the hypersurface changes under coordinate transformation in the bulk.
For this reason, it had been common to use a part of gauge freedom to fix the coordinate position of the brane.
This certainly simplifies the analysis in some cases, but it is a pity that the double covariance is lost.
In \cite{Mukohyama:2000ga} a doubly covariant perturbation scheme was given for the junction condition and it was applied to cosmological perturbation in the Randall--Sundrum braneworld.

% Codimension 2 - Purpose of this paper -
The purpose of this paper is to extend the doubly covariant perturbation scheme of Refs.~\cite{Mukohyama:2000ga,Mukohyama:2001pb} to the codimension-2 case.
As already stated, we shall restrict our attention to the situation in which the bulk geometry around the brane is axisymmetric and the surface energy--momentum tensor includes vacuum energy only.
An advantage of the doubly covariant formulation is that we have more freedom for the choice of coordinates in the bulk than other formulations without double covariance.
We shall take this advantage when we analyze the boundary condition for perturbation of six-dimensional warped flux compactification.

% Application of the formalism
As an application of our formalism, we investigate a six-dimensional braneworld model.
In this model one or two $ 3 $-branes with tension are embedded in an axisymmetric warped spacetime, and the volume of compactified two dimensions is stabilized by virtue of a magnetic flux.
The world-volume of $ 3 $-branes are $ 4 $-dimensional timelike surfaces, and one of them is considered as our universe.
Specifically, we consider geometry of the form 
\begin{align}
\begin{split}
g_{MN} \mathrm dx^M \mathrm dx^N
 & = r^2 q_{\mu\nu}(x) \mathrm dx^\mu \mathrm dx^\nu
     + \frac{\mathrm dr^2}{f(r)}
     + f(r) \mathrm d\phi^2, \\
A_M \mathrm dx^M
 & = A(r) \mathrm d\phi,
\end{split}
\label{eq:bg.gen}
\end{align}
where $ q_{\mu\nu} $ is a four-dimensional metric\footnote{
In this paper, signatures are `almost positive': $ (-+ \cdots +) $.
}.
A brane is located at $ r = r_\mathrm b $, where $ r_\mathrm b $ is a zero of $ f(r) $, and thus $ q_{\mu\nu} $ actually represents the induced metric on the brane up to the constant conformal factor $ r_\mathrm b^2 $.
The region with $ f(r) > 0 $ represents the physical domain of the extra dimensions, while the region with $ f(r) < 0 $ is unphysical and forbidden.

This model has many interesting features.
Indeed, within the ansatz (\ref{eq:bg.gen}), we can find a family of exact solutions of a braneworld with stabilized extra-dimensions.
As already mentioned, one of the most important issues in higher-dimensional theories is moduli stabilization.
For instance, in the Randall--Sundrum type-1 model, the modulus associated with the distance between the two branes acts as an additional degree of freedom appearing in low-energy description of gravity on the branes.
As a result, the ordinary four-dimensional Einstein gravity would not recover, unless this modulus, called radion, is stabilized by some non-trivial mechanisms such as a bulk scalar field with suitable potentials.
Unfortunately, no simple exact solution with a stabilized extra-dimension has been known in the context of Randall--Sundrum type, five-dimensional braneworlds.
On the other hand, with two (or more) extra-dimensions, one can introduce a $ U(1) $ gauge field (or an anti-symmetric field) to help stabilize the volume of extra dimensions.
This is because all indices of non-vanishing components of the field strength can be entirely within the extra dimensions.
The family of exact solutions is given by \cite{Gibbons:1986wg,Birmigham:1999,Gibbons:2002pq,Mukohyama:2005yw}
\begin{align}
\begin{split}
f
 & = 1 - \frac{\Lambda_6}{10} r^2 - \frac{\mu_b}{r^3} - \frac{b^2}{12 r^6}, \\
A
 & = \frac{b}{3 r^3},
\end{split}
\label{eq:bg.dS}
\end{align}
and $ q_{\mu\nu} $ is set to be the metric of a four-dimensional Einstein space with the four-dimensional effective cosmological constant $ 3 $.
Here, $ \Lambda_6 > 0 $ is the bulk cosmological constant, $ \mu_b < 0 $ is a integration constant which describes the strength of the bulk Weyl tensor, and $ b $ corresponds to the magnetic charge of the $ U(1) $ field.
The induced metric $ r_\mathrm b^2 q_{\mu\nu} $ on the brane is also an Einstein space but with the effective four-dimensional cosmological constant $ 3/r_\mathrm b^2 $.
In \cite{Mukohyama:2005yw}, the authors considered this family of solutions with $ q_{\mu\nu} $ being the de Sitter metric and showed that the behavior of gravity on the brane recovers four-dimensional one at low energies.

It is also possible to consider situations in which the effective cosmological constant of the induced metric on the brane vanishes.
By taking the limit $ r_\mathrm b \to \infty $ and properly rescaling physical quantities, the above solution becomes 
\begin{align}
\begin{split}
f
 & = - \frac{\Lambda_6}{10} r^2 - \frac{\mu_b}{r^3} - \frac{b^2}{12 r^6}, \\
A
 & = \frac{b}{3 r^3},
\end{split}
\label{eq:bg.flat}
\end{align}
and $ q_{\mu\nu} $ is the metric of a four-dimensional Einstein space with a vanishing effective cosmological constant.
In \cite{Yoshiguchi:2005nn}, it was shown that, at least when $ q_{\mu\nu} $ is the Minkowski metric, this solution is stable against linear perturbations.
Thus, the four-dimensional Einstein theory should be recovered in the linear and sufficiently low-energy regime.
In that paper, we also showed relevant Kaluza--Klein (KK) spectra of the perturbations\footnote{
In \cite{Yoshiguchi:2005nn} it was concluded that there are zero modes in the tensor sector only.
Actually, there are zero modes also in the vector sector as shown in section \ref{sec:appl.stab.vector} of the present paper.
The existence of the vector zero modes does not change the main conclusion of \cite{Yoshiguchi:2005nn} that the background is stable against linear perturbations and that the four-dimensional Einstein theory should be recovered at low energy.
}.

It is worth mentioning that these families of solutions exhibit some essential features of the warped flux compactification, including (i) warped extra-dimensions, (ii) moduli stabilization by a magnetic flux, and (iii) existence of brane(s).
The warped flux compactification has been playing important roles in the studies of moduli stabilization in the framework of string theories.
Therefore it is worthwhile to investigate these families of solutions to better understand general properties of gravity in warped flux compactification.

There is also a variety of six-dimensional braneworld models which have similar structures to ours.
Readers who are interested in such models should refer to, e.g., \cite{Papantonopoulos:2006uj} and references therein.
For other recent progresses, see also \cite{Louko:2001ik,Carter:2006uk,Lee:2006ge,Peloso:2006cq,Minamitsuji:2006cx}.

% Organization of this paper
The rest of this paper is organized as follows.
In section \ref{sec:formalism}, the formalism of the boundary condition for codimension-2 braneworlds is given.
The boundary condition is doubly covariant with respect to coordinate transformations in the bulk and on the brane.
In section \ref{sec:appl}, application of our formalism to the above six-dimensional model is presented.
In particular, it is analytically shown that in the limit where the warp factor is unity, the braneworld model is stable against linear perturbations.
We summarize the result and discuss our future works in section \ref{sec:concl}.

%%%%%%%%%%%%%%%%%%%%%%%%%%%%%%%%%%%%%%%%%%%%%%%%%%%%%%%%%%%%%%%%%%%%%%%%
\section{
Formalism
}
\label{sec:formalism}

As already explained, the main subject of this paper is the dynamics of a brane with codimension 2.
We would like to extract universal low-energy physics as much as possible without specifying microphysical properties associated with the internal structure of the brane.
For this purpose we would like to take the thin limit, i.e., the limit in which the thickness of the brane is sufficiently small compared with physical scales of interest.
This limit can be considered as just a low-energy limit, but it is known that rigorous treatment of this limit does not allow us to analyze a general setup but requires us to consider rather restricted situations.
In particular, we have to assume that there is an axisymmetry in the extra dimensions at least in a neighborhood of the world-volume of the brane, that the brane is at the center of the symmetry and that the surface energy--momentum tensor includes tension only, i.e., a component proportional to the induced metric.
Regardless of these restrictions, the thin limit has been useful for the analysis of codimension-2 objects such as strings in four-dimensions and branes in higher dimensions.

In this paper we shall still take these assumptions since we are interested in low-energy physics independent of the internal structure of the brane.
The boundary condition at the brane is conveniently described by the following well-known relation between the deficit angle $ \delta $ and the brane tension $ \sigma $:
\begin{equation}
\delta = \frac{\sigma}{M_{D+2}^D},
\label{eq:deficit-tension}
\end{equation}
where $ M_{D+2} $ is the Planck scale in the bulk spacetime.
The deficit angle $ \delta $ in the left hand side of this formula can be easily calculated if the equation determining the brane position is explicitly written as $ w = w_\mathrm b $, where $ w $ is a bulk coordinate perpendicular to the world-volume of the brane and $ w_\mathrm b $ is a constant.

However, it is not always convenient to use such a coordinate system that one of the bulk coordinates is perpendicular to the world-volume of the brane and that the brane position is specified by a fixed value of this radial coordinate.
Actually, in a coordinate system natural and/or convenient for the analysis of the bulk geometry, the brane position is often specified not by a fixed coordinate value but by parametric equations.
In some cases including the examples in section \ref{sec:appl}, the parametric equations even involve unknown functions, which are determined only after solving some dynamical equations.
Therefore it would be more convenient if there were a simple way to calculate the deficit angle in general bulk coordinates.

In the doubly-covariant formalism developed in this section, the deficit angle can be calculated directly in any coordinate system and, thus, it allows more freedom for the choice of bulk coordinates.
We shall take the full advantage of this freedom when we seek the boundary condition for perturbations of six-dimensional brane-worlds in section \ref{sec:appl}.

%%%%%%%%%%%%%%%%%%%%%%%%%%%%%%%%%%%%%%%%%%%%%%%%%%%%%%%%%%%%%%%%%%%%%%%%
\subsection{
Deficit angle and boundary condition for axisymmetric configurations
}
\label{sec:formalism.bc}

Let us consider a $ (D+2) $-dimensional bulk spacetime $ (\mcm,g) $ and a $ D $-dimensional world-volume $ \mcb $ of a $ (D-1) $-brane embedded in it\footnote{
A $ p $-brane is a $ p $-dimensional extended object and, thus, its world-volume is a $ (p+1) $-dimensional timelike surface.
}.
The embedding is specified by the set of functions $ Z = \{Z^M(y)\} $ as 
\begin{equation}
x^M = Z^M(y),
\end{equation}
where $ x = \{ x^M\} $ denotes coordinates in the bulk and $ y = \{ y^{\mu}\} $ denotes coordinates on the world-volume of the brane.
Hereafter, capital Latin indices run over all the dimensions $ M,N,\ldots = 0,\ldots,D+1 $, and Greek and small Latin indices run, respectively, over the brane and the extra dimensions, i.e., $ \mu,\nu,\ldots = 0,\ldots,D-1 $ and $ a,b,\ldots = D,D+1 $.

A neighborhood of $ \mcb $ can be foliated by a $ 2 $-parameter family of $ D $-dimensional timelike surfaces.
Let us denote the corresponding embeddings by $ x^M = {Z_\zeta}^M(y) $, where $ \zeta = \{\zeta_1,\zeta_2\} $ parameterizes the $ D $-dimensional timelike surfaces, and suppose that $ {Z_{\zeta=0}}^M(y) = Z^M(y) $.
The foliation is necessary just to make the geometrical quantities such as $ {e_\mu}^M $ and $ q_{\mu\nu} $ (see below for their definitions) well-defined off the brane.
The final expression of the boundary condition does not depend on the way the neighborhood of $ \mcb $ is foliated, and is written in terms of local quantities on the brane only.

Note that the coordinate system on the world-volume of the brane does not need to be a subset of that in the bulk.
Thus, the coordinate transformations
\begin{equation}
x^M
 \to {x'}^M = F^M(x), \quad
y^\mu
 \to {y'}^\mu = f^\mu(y)
\end{equation}
in the bulk and on the world-volume of the brane, respectively, are disentangled and are independently defined.
Since the physical position of the brane is independent of the choice of coordinates, the embedding functions transform respectively as 
\begin{equation}
{Z_\zeta}^M(y)
 \to {Z_\zeta'}^M(y) = F^M(Z_\zeta(y)), \quad
{Z_\zeta}^M(y)
 \to {Z_\zeta''}^M(y) = Z_\zeta(f^{-1}(y))
\end{equation}
under these transformations, where $ f^{-1} $ is the inverse of the map $ f: y^{\mu} \mapsto f^{\mu}(y) $.
Because of this property, the tangent vectors $ {e_\mu}^M $ defined by 
\begin{equation}
{e_\mu}^M
 \equiv \frac{\partial {Z_\zeta}^M}{\partial y^\mu}
\label{eq:e.def}
\end{equation}
transform as vectors under the bulk coordinate transformation.
As a consequence, the product $ {e_\mu}^M {e_\nu}^N g_{MN} $ transforms as a scalar under the bulk coordinate transformation and, thus, the induced metric
\begin{equation}
q_{\mu\nu}
 \equiv \left.{e_\mu}^M {e_\nu}^N g_{MN}\right|_{x=Z(y)}
\end{equation}
is actually invariant under the bulk coordinate transformation.
On the other hand, under the coordinate transformation on the world-volume of the brane, the induced metric $ q_{\mu\nu} $ does indeed transform as a tensor.

Now we assume that the two-dimensional extra space is axisymmetric and introduce the corresponding spacelike Killing vector field $ \varphi^M $ in the $ (D+2) $-dimensional bulk spacetime and angular coordinate $ \phi $.
They satisfy $ \varphi^M = (\partial/\partial\phi)^M $ and $ \nabla_M \varphi_N + \nabla_N \varphi_M = 0 $, where $ \nabla_M $ is the covariant derivative associated with $ g_{MN} $.
We suppose that a brane is placed at the axis of the axisymmetry.
Namely, we impose the condition
\begin{equation}
\left.L\right|_{x=Z(y)} = 0,
\label{eq:L}
\end{equation}
where
\begin{equation}
L
 \equiv \sqrt{g_{MN} \varphi^M \varphi^N}.
\label{eq:L.def}
\end{equation}
In the beginning of this subsection we introduced a $ 2 $-parameter family of $ D $-dimensional timelike surfaces $ x^M = {Z_\zeta}^M(y) $ to foliate a neighborhood of the world-volume $ \mcb $ of the brane.
We suppose that this also respects the axisymmetry, i.e., each surface in the family is orthogonal to $ \varphi $:
\begin{equation}
g_{MN} {e_\mu}^M \varphi^N = 0.
\end{equation}

We shall relate the deficit angle to the normal derivative of the scalar $ L $.
For this purpose we introduce a unit vector field $ n^M $ normal to the surface $ x^M = {Z_\zeta}^M(y) $ and the Killing vector $ \varphi^M $:
\begin{align}
\begin{split}
& n_M \varphi^M = n_M {e_\mu}^M = 0, \\
& n_M n^M = 1.
\end{split}
\end{align}
These conditions determine $ n^M $ up to the overall sign, which is fixed by demanding that $ n^M $ should direct from the brane to the bulk.
Note that $ n^M $ is well-defined not only on the world-volume $ \mcb $ of the brane but also in a neighborhood of it.
The deficit angle $ \delta $ can now be written in a doubly-covariant way in terms of the embedding functions $ Z^M(y) $, the unit norm $ n^M $ and the scalar $ L $:
\begin{equation}
\left.\partial_\perp L\right|_{x=Z(y)}
 = \frac{2\pi-\delta}{\Delta\phi},
\label{eq:dL}
\end{equation}
where $ \partial_\perp = n^M \partial_M $ is the normal derivative and $ \Delta\phi $ is the period of the angular coordinate $ \phi $, i.e., $ \phi \sim \phi + \Delta\phi $.

In summary, the boundary condition at the brane is given by (\ref{eq:L}) and (\ref{eq:deficit-tension}), where the deficit angle $ \delta $ is given by the doubly-covariant formula (\ref{eq:dL}).

The doubly covariant boundary condition written above in terms of $ L $ is easy to calculate.
Actually, we show an explicit result for a concrete example of background 6D warped flux compactification in section \ref{sec:appl}.
However, when we perform perturbative analysis around some background and seek the corresponding perturbative expansion of the boundary condition, this form is not most convenient since the definition (\ref{eq:L.def}) of $ L $ involves a square-root and thus is not an analytic function of metric components.
Therefore, for the purpose of perturbative analysis, it is more convenient to use the following alternative expression:
\begin{align}
\mcl_0
 & \equiv \left.L^2\right|_{x=Z(y)} = 0,
\label{eq:L0}\\
\mcl_1
 & \equiv \left.\partial_\perp L^2\right|_{x=Z(y)} = 0,
\label{eq:L1}\\
\mcl_2
 & \equiv \left.\partial_\perp^2 L^2\right|_{x=Z(y)}
   = 2 \left(\frac{2\pi-\delta}{\Delta\phi}\right)^2,
\label{eq:L2}
\end{align}
where $ \delta $ is related to the brane tension $ \sigma $ by (\ref{eq:deficit-tension}).
It is evident that $ \mcl_{0,1,2} $ are analytic functions of the metric components.
Therefore, it is easy to perform perturbative expansion.

One must, however, be aware of the fact that it is only up to overall sign of $ \partial_\perp L|_{x=Z(y)} $ that this form of the boundary condition is equivalent to the original boundary condition.
Actually this does not matter if $ \mcl_2 $ does not vanish for the background and if perturbation is small enough in the sense that the background value of $ 2 \pi - \delta = 2 \pi - \sigma/M_{D+2}^D $ and the corresponding perturbed value have the same sign.
This is the case unless the background value $ \sigma $ is fine-tuned to the special value $ 2 \pi M_{D+2}^D $ with an extremely high accuracy.
Anyway, if $ \sigma $ had such a large value ($ \sim 2 \pi M_{D+2}^D $) then the effective low-energy description would break down.
Thus, while we can apply the original form of the boundary condition ((\ref{eq:L}) and (\ref{eq:dL}) with (\ref{eq:deficit-tension})) to the background, we can use the alternative form ((\ref{eq:L0}), (\ref{eq:L1}) and (\ref{eq:L2}) with (\ref{eq:deficit-tension})) for perturbation.

%%%%%%%%%%%%%%%%%%%%%%%%%%%%%%%%%%%%%%%%%%%%%%%%%%%%%%%%%%%%%%%%%%%%%%%%
\subsection{
Linear perturbation
}

The formalism developed in the previous subsection displays its real ability when used in perturbative analysis.
Since the doubly-covariant boundary condition is written in terms of scalar quantities which are analytic functions of metric components, it is manifest how to expand it with respect to the perturbation.
In this subsection we consider metric perturbation and perturbation of the embedding functions (representing the position of the brane), and expand the boundary condition up to the linearized order.

We still assume axisymmetry and, thus, we consider axisymmetric perturbations only.

%%%%%%%%%%%%%%%%%%%%%%%%%%%%%%%%%%%%%%%%%%%%%%%%%%%%%%%%%%%%%%%%%%%%%%%%
\subsubsection{
Perturbed boundary condition
}
\label{sec:formalism.perturb.bc}

Here we consider linearly perturbed geometry.
As has been stressed, our purpose is to formulate everything in a manner which is covariant both in the bulk and on the brane.
Since the coordinate transformation in the bulk changes the coordinate position of the brane, the double covariance requires us to consider perturbation of the brane position as well as metric perturbation.
The most general perturbations of the background geometry and the position of the brane are 
\begin{align}
\begin{split}
& g_{MN} = g^{(0)}{}_{MN} + \delta g_{MN}, \quad
  \varphi^M = \varphi^{(0)}{}^M + \delta\varphi^M, \\
& Z^M(y) = Z^{(0)}{}^M(y) + \delta Z^M(y),
\end{split}
\label{eq:perturb.geom.def}
\end{align}
where, throughout this paper, background quantities are recognized by the script `$ (0) $'.
Other linearly perturbed quantities are calculated in terms of $ \delta g_{MN} $, $ \delta\varphi^M $ and $ \delta Z^M $, and indices of perturbed tensors are raised and lowered by $ g^{(0)}{}^{MN} $ and $ g^{(0)}{}_{MN} $, respectively.
Note, for example, that $ g^{MN} = g^{(0)}{}^{MN} - \delta g^{MN} $ up to the linear order.

In the following we shall express the linear perturbations of $ \mcl_{0,1,2} $ in terms of $ \delta g_{MN} $, $ \delta\varphi^M $ and $ \delta Z^M $, where $ \mcl_{0,1,2} $ are defined by (\ref{eq:L0}), (\ref{eq:L1}), and (\ref{eq:L2}).

Before proceeding, we here make a comment on the perturbed Killing vector $ \varphi^M $.
It is always possible to set $ \delta\varphi^M = 0 $ by using the gauge freedom and to set $ \delta(\Delta\phi) = 0 $ by rescaling the angular coordinate $ \phi $.
In the following analysis, however, while we shall set $ \delta(\Delta\phi) = 0 $, we shall suppose that $ \delta\varphi^M $ is not necessarily vanishing.
This is because, in order to explicitly confirm that our perturbative scheme maintains the double covariance, we need to include $ \delta\varphi^M $ and its gauge transformation.
Thus, we keep $ \delta\varphi^M $ until section \ref{sec:formalism.perturb.double_covariance}, where the perturbative double covariance is confirmed, but always set $ \delta(\Delta\phi) = 0 $.
The perturbation $ \delta\varphi^M $ of the Killing vector should satisfy the linearized Killing equation 
\begin{align}
0
 & = \delta\left(\nabla_M \varphi_N + \nabla_N \varphi_M\right) \nonumber\\
 & = -2 \delta\Gamma_{MN}^P \varphi^{(0)}{}_P
     + \nabla^{(0)}{}_M \delta\varphi_N
     + \nabla^{(0)}{}_N \delta\varphi_M
     + \nabla^{(0)}{}_M \varphi^{(0)}{}^P \delta g_{PN}
     + \nabla^{(0)}{}_N \varphi^{(0)}{}^P \delta g_{PM},
\end{align}
and should be orthogonal to the surface $ x^M = Z^M(y) $ so that
\begin{equation}
0
 = \delta\left(\delta g_{MN} \varphi^M {e_\mu}^N\right)
 = \delta g_{MN} \varphi^{(0)}{}^M e_\mu^{(0)}{}^N
   + \delta\varphi_M e_\mu^{(0)}{}^M
   + \varphi^{(0)}{}_M \delta{e_\mu}^M,
\end{equation}
where $ \delta\Gamma^P_{MN} \equiv g^{(0)}{}^{PQ} (\nabla^{(0)}{}_N \delta g_{QM} + \nabla^{(0)}{}_M \delta g_{NQ} - \nabla^{(0)}{}_Q \delta g_{MN})/2 $ is the linear perturbation of the Christoffel symbol, and $ \delta{e_\mu}^M $ is that of the tangent vector:
\begin{equation}
{e_\mu}^M
 = e_\mu^{(0)}{}^M + \delta{e_\mu}^M.
\label{eq:deltae.def}
\end{equation}
(We shall give a concrete expression for $ \delta{e_\mu}^M $ in (\ref{eq:deltae}).) 
Needless to say, when we apply our formalism to concrete problems, it is more convenient to set $ \delta\varphi^M = 0 $ by a gauge transformation than keeping it non-vanishing throughout.
Therefore, in section \ref{sec:formalism.perturb.simple} we shall simplify the perturbed boundary condition by setting $ \delta\varphi^M = 0 $ for later convenience.
In section \ref{sec:appl}, when we analyze perturbations of a particular six-dimensional braneworld model, we shall use the simplified expression.

Let us now go back to the formulation of the perturbative scheme.

First, we calculate the linear perturbation of the induced metric.
The perturbation $ \delta q_{\mu\nu} $ is defined in the following manifest way:
\begin{equation}
q_{\mu\nu}(y) = q_{\mu\nu}^{(0)}(y) + \delta q_{\mu\nu}(y),
\end{equation}
where, up to the linear order,
\begin{equation}
q_{\mu\nu}(y)
 \equiv \left.g_{MN} {e_\mu}^M {e_\nu}^N\right|_{x=Z(y)}
 = \Bigl.g_{MN} {e_\mu}^M {e_\nu}^N\Bigr|_{x=Z^{(0)}(y)}
   + \Bigl.
     \delta Z^L \partial_L
     \left(g^{(0)}{}_{MN} e_\mu^{(0)}{}^M e_\nu^{(0)}{}^N\right)
     \Bigr|_{x=Z^{(0)}(y)}
\end{equation}
and
\begin{equation}
q_{\mu\nu}^{(0)}(y)
 \equiv \left.g^{(0)}{}_{MN} e_\mu^{(0)}{}^M e_\nu^{(0)}{}^N\right|_{x=Z^{(0)}(y)}.
\end{equation}
Here, in order to get the last expression we have used the fact that $ g_{MN} {e_\mu}^M {e_\nu}^N $ is a scalar in the bulk and kept only terms up to the linear order.
Thus we obtain
\begin{equation}
\delta q_{\mu\nu}(y)
 = \Bigl.\delta\left(g_{MN} {e_\mu}^M {e_\nu}^N\right)\Bigr|_{x=Z^{(0)}(y)}
   + \Bigl.
     \delta Z^L\partial_L
     \left(g^{(0)}{}_{MN} e_\mu^{(0)}{}^M e_\nu^{(0)}{}^N\right)
     \Bigr|_{x=Z^{(0)}(y)}.
\label{eq:deltaq.pre}
\end{equation}
In general, linear perturbation of a quantity of the form $ X|_{x=Z(y)} $ is evaluated as 
\begin{equation}
\delta\left(\left.X\right|_{x=Z(y)}\right) 
 = \delta\Bigl.X\Bigr|_{x=Z^{(0)}(y)}
   + \Bigl.\delta Z^L\partial_L X^{(0)}\Bigr|_{x=Z^{(0)}(y)},
\label{eq:deltaX.def}
\end{equation}
where $ X $ is a scalar in the bulk.

In order to evaluate the first term in the right hand side of (\ref{eq:deltaq.pre}), we need to express $ \delta{e_\mu}^M $.
Noting that $ {e_\mu}^M $ and $ e_\mu^{(0)}{}^M $ in equation (\ref{eq:deltae.def}) should be at a common point (say, at $ x = Z $ or $ x = Z^{(0)} $), it is understood that the linear perturbation $ \delta{e_\mu}^M $ can be written in terms of the Lie derivative as
\begin{equation}
\delta{e_\mu}^M
 = \Lie_{e_\mu^{(0)}} \delta Z^M
 = -\Lie_{\delta Z} e_\mu^{(0)}{}^M,
\label{eq:deltae}
\end{equation}
where $ \Lie $ denotes the Lie derivative in the bulk.
(See equation (\ref{eq:e.def}) for the definition of $ {e_\mu}^M $.)
This expression allows us to calculate the linear perturbation of the induced metric.
Combining (\ref{eq:deltae}) and (\ref{eq:deltaq.pre}), we can calculate $ \delta q_{\mu\nu}(y) $ as
\begin{align}
\delta q_{\mu\nu}(y)
 & = \left[
     \left(
     \delta g_{MN} e_\mu^{(0)}{}^M e_\nu^{(0)}{}^N
     + g^{(0)}{}_{MN} \delta{e_\mu}^M e_\nu^{(0)}{}^N
     + g^{(0)}{}_{MN} e_\mu^{(0)}{}^M \delta{e_\nu}^N
     \right)
     + \delta Z^L \partial_L
       \left(g^{(0)}{}_{MN} e_\mu^{(0)}{}^M e_\nu^{(0)}{}^N\right)
     \right]_{x=Z^{(0)}(y)} \nonumber\\
 & = \left.\left(\delta g_{MN} + \Lie_{\delta Z} g^{(0)}{}_{MN}\right)
           e_\mu^{(0)}{}^M e_\nu^{(0)}{}^N\right|_{x=Z^{(0)}(y)}.
\label{eq:deltaq.def}
\end{align}

Second, let us calculate the linear perturbation of $ \mcl_0 $ defined in (\ref{eq:L0}).
By the aid of the general formula (\ref{eq:deltaX.def}) with $ X = g_{MN} \varphi^M \varphi^N $, this is an easy task.
\begin{align}
\delta\mcl_0(y)
 & = \delta\left(\left.g_{MN}\varphi^M\varphi^N\right|_{x=Z(y)}\right) \nonumber\\
 & = \left[\delta\left(g_{MN}\varphi^M\varphi^N\right)
     + \delta Z^L\partial_L
       \left(g^{(0)}{}_{MN}\varphi^{(0)}{}^M\varphi^{(0)}{}^N\right)
     \right]_{x=Z^{(0)}(y)} \nonumber\\
 & = \left[2 \varphi^{(0)}{}_M \delta\varphi^M
     + \varphi^{(0)}{}^M \varphi^{(0)}{}^N \delta g_{MN}
     + \delta Z^L \partial_L
       \left(\varphi^{(0)}{}^P \varphi^{(0)}{}_P\right)
     \right]_{x=Z^{(0)}(y)}.
\label{eq:deltaL0}
\end{align}

Third, let us calculate the linear perturbation of $ \mcl_1 $ defined in (\ref{eq:L1}).
By applying the formula (\ref{eq:deltaX.def}) to $ X = \partial_\perp (g_{MN} \varphi^M \varphi^N) $, we obtain
\begin{align}
\delta\mcl_1(y)
 & = \delta\left(
     \left.\partial_\perp \left(g_{MN} \varphi^M \varphi^N\right)\right|_{x=Z(y)}
     \right) \nonumber\\
 & = \left[\delta\left(
           \partial_\perp \left(g_{MN} \varphi^M \varphi^N\right)
           \right)
           + \delta Z^L\partial_L \left(\partial_\perp^{(0)}
             \left(g^{(0)}{}_{MN} \varphi^{(0)}{}^M \varphi^{(0)}{}^N\right)
             \right)
     \right]_{x=Z^{(0)}(y)}.
\label{eq:deltaL1.pre}
\end{align}
To proceed further, we need to calculate the perturbation of the normal derivative $ \partial_\perp = g^{MN} n_M \partial_N $.
The perturbation of $ n_M $ is actually determined by the following orthonormality condition:
\begin{align}
\begin{split}
0 & = \delta\left(n_M \varphi^M\right)
    = \delta n_M \varphi^{(0)}{}^M + n^{(0)}{}_M \delta\varphi^M, \\
0 & = \delta\left(n_M {e_\mu}^M\right)
    = \delta n_M e_\mu^{(0)}{}^M - n^{(0)}{}_M \Lie_{\delta Z} e_\mu^{(0)}{}^M, \\
0 & = \delta\left(n_M n^M\right)
    = 2 \delta n_M n^{(0)}{}^M - n^{(0)}{}^M n^{(0)}{}^N \delta g_{MN},
\end{split}
\end{align}
where we have defined $ \delta n_M $ as the linear perturbation of $ n_M $ with a lower index: $ n_M = n^{(0)}{}_M + \delta n_M $.
From these conditions, $ \delta n_M $ is uniquely decomposed as
\begin{equation}
\delta n_M
 = \frac{1}{2} n^{(0)}{}^P n^{(0)}{}^Q \delta g_{PQ} n^{(0)}{}_M
   - \frac{n^{(0)}{}_P \delta\varphi^P}{\varphi^{(0)}{}^Q \varphi^{(0)}{}_Q}
     \varphi^{(0)}{}_M
   + q^{(0)\mu\nu} n^{(0)}{}_N \Lie_{\delta Z} e_\nu^{(0)}{}^N e_\mu^{(0)}{}_M.
\end{equation}
Remember again that this is the perturbation of $ n_M $ with the lower index and that just raising the index of this expression by $ g^{(0)}{}^{M'M} $ does not give the correct formula for the perturbation of $ n^{M'} $.
Indeed, the perturbation of $ n^M $ is given by $ \delta(n^M) = \delta n^M - \delta g^{MN} n^{(0)}{}_N $ and, thus,
\begin{equation}
\delta(\partial_\perp)
 = \left(\delta n^M - \delta g^{MN} n^{(0)}{}_N\right) \partial_M.
\end{equation}
Using this expression for $ \delta(\partial_\perp) $, equation (\ref{eq:deltaL1.pre}) gives the following expression for $ \delta\mcl_1 $:
\begin{multline}
\delta\mcl_1(y)
 = \Bigl[
   \left(\delta n^M - n^{(0)}{}_N \delta g^{MN} \right) \partial_M
   \left(\varphi^{(0)}{}^P \varphi^{(0)}{}_P\right)
   + n^{(0)}{}^M \partial_M
     \left(2 \varphi^{(0)}{}_P \delta\varphi^P
           + \varphi^{(0)}{}^P \varphi^{(0)}{}^Q \delta g_{PQ}\right)
   \Bigr. \\
   \Bigl.
   + \delta Z^L \partial_L
     \left(n^{(0)}{}^M \partial_M
           \left(\varphi^{(0)}{}^P \varphi^{(0)}{}_P\right)\right)
   \Bigr]_{x=Z^{(0)}(y)}.
\label{eq:deltaL1}
\end{multline}

Fourth, perturbation of $ \mcl_2 $ defined in (\ref{eq:L2}) is also calculated in a similar way as
\begin{align}
\delta\mcl_2(y)
 & = \delta\left(\left.\partial_\perp^2 L^2\right|_{x=Z(y)}\right) \nonumber\\ 
 & = \biggl[
     \left(\delta n^M - n^{(0)}{}_N \delta g^{MN}\right) \partial_M
     \left(n^{(0)}{}^L \partial_L
           \left(\varphi^{(0)}{}^P \varphi^{(0)}{}_P\right)\right)
     + n^{(0)}{}^L \partial_L
       \left(\left(\delta n^M - n^{(0)}{}_N \delta g^{MN} \right) \partial_M
             \left(\varphi^{(0)}{}^P \varphi^{(0)}{}_P\right) \right)
     \biggr. \nonumber\\
 &   \qquad
     \biggl.
     + \left(n^{(0)}{}^M \partial_M\right)^2
       \left(2 \varphi^{(0)}{}_P \delta\varphi^P
             + \varphi^{(0)}{}^P \varphi^{(0)}{}^Q \delta g_{PQ}\right)
     + \delta Z^L \partial_L
       \left(\left(n^{(0)}{}^M \partial_M\right)^2
             \left(\varphi^{(0)}{}^P \varphi^{(0)}{}_P\right)\right)
     \biggr]_{x=Z^{(0)}(y)}.
\label{eq:deltaL2}
\end{align}

Finally, the perturbed boundary condition is obtained as
\begin{equation}
\delta\mcl_0 = \delta\mcl_1 = \delta\mcl_2 = 0,
\end{equation}
where $ \delta\mcl_{0,1,2} $ are given by (\ref{eq:deltaL0}), (\ref{eq:deltaL1}), and (\ref{eq:deltaL2}), respectively.

%%%%%%%%%%%%%%%%%%%%%%%%%%%%%%%%%%%%%%%%%%%%%%%%%%%%%%%%%%%%%%%%%%%%%%%%
\subsubsection{
Boundary condition for $ U(1) $ gauge potential
}

If a $ U(1) $ gauge potential $ A_M \mathrm dx^M $ is present in the bulk, one must impose a separate boundary condition on the $ U(1) $ field.
Here the perturbed $ U(1) $ gauge potential is denoted as 
\begin{equation}
A_M = A^{(0)}{}_M + \delta A_M.
\label{eq:perturb.gauge.def}
\end{equation}

When we analyze physical perturbation of a background configuration, one should keep the total magnetic flux unchanged since it is a conserved quantity.
With the total magnetic flux being the same as the background value, one can set the perturbation of $ \mca \equiv A_M \varphi^M|_{x=Z(y)} $ to zero on all the boundaries for the following reasons.
First, the $ y $-dependence of $ \mca $ is forbidden by the regularity of $ F^{MN} F_{MN} $.
Second, the $ y $-independent perturbation of $ \mca $ is not forbidden by the regularity but can be gauged away on all boundaries simultaneously since the total magnetic flux is assumed to be the same as the background.
Therefore, we can set the perturbation of $ \mca $ to be zero on all boundaries.

To calculate the perturbation of this quantity, one must take into account the fact that the position $ Z $ of the brane is also perturbed in general.
Thus, the boundary condition for the perturbation of the $ U(1) $ field is 
\begin{equation}
\delta\mca = 0,
\end{equation}
where 
\begin{align}
\delta\mca(y)
 & \equiv \delta\left(\left.A_M \varphi^M\right|_{x=Z(y)}\right) \nonumber\\
 & = \left[
     \delta\left(A_M \varphi^M\right)
       + \delta Z^L\partial_L
         \left(A^{(0)}{}_M \varphi^{(0)}{}^M\right)
     \right]_{x=Z^{(0)}(y)} \nonumber\\
 & = \left[
     \delta A_M \varphi^{(0)}{}^M
     + A^{(0)}{}_M \delta\varphi^M
     + \delta Z^L \partial_L
       \left(A^{(0)}{}_M \varphi^{(0)}{}^M\right)
     \right]_{x=Z^{(0)}(y)}.
\label{eq:deltaA}
\end{align}
Here, we have used the formula (\ref{eq:deltaX.def}) with $ X = A_M \varphi^M $ to get the second line.

Note that, in general, it is not possible to set $ A_M \varphi^M $ itself to zero on more than one boundaries simultaneously.
Indeed, with a non-zero total magnetic flux, a gauge in which $ A_M \varphi^M $ is zero on a boundary is not compatible with another gauge in which $ A_M \varphi^M $ is zero on another boundary.
The total magnetic flux, which is gauge-invariant, can indeed be expressed in terms of the difference between values of $ A_M \varphi^M $ at different boundaries;
it is not the value of $ A_M \varphi^M|_{x=Z(y)} $ itself but the difference $ \delta\mca $ that can be set to zero on all branes simultaneously.

%%%%%%%%%%%%%%%%%%%%%%%%%%%%%%%%%%%%%%%%%%%%%%%%%%%%%%%%%%%%%%%%%%%%%%%%
\subsubsection{
Double covariance
}
\label{sec:formalism.perturb.double_covariance}

The general formalism developed in section \ref{sec:formalism.bc} is doubly covariant in the sense that it is covariant under coordinate transformation on the brane and invariant under coordinate transformation in the bulk.
Therefore, the corresponding boundary conditions for linear perturbations should have double covariance up to the linearized order.
In the following we shall confirm this explicitly as a consistency check.

What we would like to show as a consistency check is that the perturbed boundary condition is doubly covariant under infinitesimal coordinate transformations in the bulk and on the brane, where the two kinds of transformation are denoted as 
\begin{align}
x^M
 & \to x^M + \xi^M(x),
\label{eq:gauge.bulk}\\
y^\mu
 & \to y^\mu + \zeta^\mu(y),
\label{eq:gauge.brane}
\end{align}
respectively.

Perturbations in the bulk, (\ref{eq:perturb.geom.def}) and (\ref{eq:perturb.gauge.def}), transform under (\ref{eq:gauge.bulk}) as 
\begin{align}
\begin{split}
\delta g_{MN}
 & \to \delta g_{MN} - \Lie_\xi g^{(0)}{}_{MN}, \\
\delta\varphi^M
 & \to \delta\varphi^M - \Lie_\xi \varphi^{(0)}{}^M, \\
\delta A_M
 & \to \delta A_M - \Lie_\xi A^{(0)}{}_M, \\
\delta Z^M
 & \to \delta Z^M + \xi^M.
\end{split}
\end{align}
Thus, it immediately follows from the formula (\ref{eq:deltaL0}) that
\begin{align}
\delta\mcl_0
 & \to \left[
       2 \varphi^{(0)}{}_M 
       \left(\delta\varphi^M - \Lie_\xi \varphi^{(0)}{}^M\right)
       + \varphi^{(0)}{}^M \varphi^{(0)}{}^N
         \left(\delta g_{MN} - \Lie_\xi g^{(0)}{}_{MN}\right)
       + \left(\delta Z^L + \xi^L\right) \partial_L
         \left(\varphi^{(0)}{}^P \varphi^{(0)}{}_P\right)
       \right]_{x=Z^{(0)}(y)} \nonumber\\
 & = \left[
     2 \varphi^{(0)}{}_M \delta\varphi^M
       + \varphi^{(0)}{}^M \varphi^{(0)}{}^N \delta g_{MN}
       + \delta Z^L \partial_L
         \left(\varphi^{(0)}{}^P\varphi^{(0)}{}_P\right)
     \right]_{x=Z^{(0)}(y)} \nonumber\\
 & = \delta\mcl_0.
\end{align}
This says that the boundary condition $ \delta\mcl_0 = 0 $ is invariant under the gauge transformation in the bulk (\ref{eq:gauge.bulk}).
Similarly, we can show that the other boundary conditions $ \delta\mcl_{1,2} = 0 $ and $ \delta\mca = 0 $ are invariant under (\ref{eq:gauge.bulk}).

On the other hand, under the gauge transformation on the brane (\ref{eq:gauge.brane}),
\begin{align}
\begin{split}
\delta g_{MN}|_{x=Z^{(0)}(y)}
 & \to \delta g_{MN}|_{x=Z^{(0)}(y)}, \\
\delta\varphi^M|_{x=Z^{(0)}(y)}
 & \to \delta\varphi^M|_{x=Z^{(0)}(y)}, \\
\delta A_M|_{x=Z^{(0)}(y)}
 & \to \delta A_M|_{x=Z^{(0)}(y)}, \\
\delta Z^M
 & \to \delta Z^M - \zeta^\mu e_\mu^{(0)}{}^M.
\end{split}
\end{align}
Then, from the formula (\ref{eq:deltaq.def}), we see that the perturbation of the induced metric on the brane transforms as 
\begin{align}
\delta q_{\mu\nu}
 & \to \left.
       \left(\delta g_{MN}
             + \left(\Lie_{\delta Z}
                     - \Lie_{\zeta^\mu e_\mu^{(0)}}\right) g^{(0)}{}_{MN}\right)
       e_\mu^{(0)}{}^M e_\nu^{(0)}{}^N
       \right|_{x=Z^{(0)}(y)} \nonumber\\
 & = \left.
     \left(\delta g_{MN} + \Lie_{\delta Z} g^{(0)}{}_{MN}\right)
     e_\mu^{(0)}{}^M e_\nu^{(0)}{}^N
     \right|_{x=Z^{(0)}(y)}
     - \Lie^D_\zeta q_{\mu\nu}^{(0)} \nonumber\\
 & = \delta q_{\mu\nu} - \Lie^D_\zeta q_{\mu\nu}^{(0)},
\end{align}
where $ \Lie^D $ is the $ D $-dimensional Lie derivative defined on the brane.
Similarly, it is shown that $ \delta\mcl_{0,1,2} $ and $ \delta\mca $ transform under (\ref{eq:gauge.brane}) as
\begin{align}
\begin{split}
\delta\mcl_0
 & \to \delta\mcl_0 - \Lie^D_{\zeta}\mcl_0^{(0)}, \\
\delta\mcl_1 
 & \to \delta\mcl_1 - \Lie^D_{\zeta}\mcl_1^{(0)}, \\
\delta\mcl_2 
 & \to \delta\mcl_2 - \Lie^D_{\zeta}\mcl_2^{(0)}, \\
\delta\mca
 & \to \delta\mca - \Lie^D_{\zeta}\mca^{(0)}.
\end{split}
\end{align}
Therefore, all equations in the perturbed boundary condition have the double covariance as they should.

%%%%%%%%%%%%%%%%%%%%%%%%%%%%%%%%%%%%%%%%%%%%%%%%%%%%%%%%%%%%%%%%%%%%%%%%
\subsubsection{
Simplified expression
}
\label{sec:formalism.perturb.simple}

We have developed a perturbation scheme for our doubly covariant boundary condition.
It is given by 
\begin{equation}
\delta\mcl_0 = \delta\mcl_1 = \delta\mcl_2 = \delta\mca = 0,
\label{eq:bc.perturb}
\end{equation}
where $ \delta\mcl_{0,1,2} $ and $ \delta\mca $ are given by (\ref{eq:deltaL0}), (\ref{eq:deltaL1}), (\ref{eq:deltaL2}), and (\ref{eq:deltaA}).
As a consistency check, we have explicitly seen that they are indeed doubly covariant in the linearized level.

So far, we have been keeping $ \delta\varphi^M $ since we had to take into account its gauge transformation in order to show the perturbative double covariance in section \ref{sec:formalism.perturb.double_covariance}.
On the other hand, when we apply our formalism to concrete problems, it is more convenient to set $ \delta\varphi^M = 0 $ by a gauge transformation than keeping it non-vanishing throughout.
With $ \delta\varphi^M = 0 $ (and $ \delta(\Delta\phi) = 0 $), we have slightly simplified expressions for $ \delta\mcl_{0,1,2} $ and $ \delta\mca $ as follows.
\begin{align}
\delta\mcl_0
 & = \left[\varphi^{(0)}{}^M \varphi^{(0)}{}^N \delta g_{MN}
     + \delta Z^L \partial_L
       \left(\varphi^{(0)}{}^P \varphi^{(0)}{}_P\right)\right]_{x=Z^{(0)}(y)}, 
\label{eq:deltaL0.simple}\\
\delta\mcl_1
 & = \Bigl[
     \left(\delta n^M - n^{(0)}{}_N \delta g^{MN} \right) \partial_M
     \left(\varphi^{(0)}{}^P \varphi^{(0)}{}_P\right)
       + n^{(0)}{}^M \partial_M
         \left(\varphi^{(0)}{}^P \varphi^{(0)}{}^Q \delta g_{PQ}\right)
     \Bigr. \nonumber\\
 &   \qquad
     \Bigl.
     + \delta Z^L \partial_L
       \left(n^{(0)}{}^M \partial_M
             \left(\varphi^{(0)}{}^P \varphi^{(0)}{}_P\right)\right)
     \Bigr]_{x=Z^{(0)}(y)}, 
\label{eq:deltaL1.simple}\\
\delta\mcl_2
 & = \biggl[
     \left(\delta n^M - n^{(0)}{}_N \delta g^{MN} \right) \partial_M
     \left(n^{(0)}{}^L \partial_L
           \left(\varphi^{(0)}{}^P \varphi^{(0)}{}_P\right)\right)
     + n^{(0)}{}^L \partial_L
       \left(\left(\delta n^M - n^{(0)}{}_N \delta g^{MN} \right) \partial_M
             \left(\varphi^{(0)}{}^P \varphi^{(0)}{}_P\right)\right)
     \biggr. \nonumber\\
 &   \qquad
     \biggl.
     + \left(n^{(0)}{}^M \partial_M\right)^2
       \left(\varphi^{(0)}{}^P \varphi^{(0)}{}^Q \delta g_{PQ}\right)
     + \delta Z^L \partial_L
       \left(\left(n^{(0)}{}^M \partial_M\right)^2
             \left(\varphi^{(0)}{}^P \varphi^{(0)}{}_P\right)\right)
     \biggr]_{x=Z^{(0)}(y)}, 
\label{eq:deltaL2.simple}\\
\delta\mca
 & = \left[
     \delta A_M \varphi^{(0)}{}^M
     + \delta Z^L \partial_L \left(A^{(0)}{}_M \varphi^{(0)}{}^M\right)
     \right]_{x=Z^{(0)}(y)}.
\label{eq:deltaA.simple}
\end{align}
The perturbation of the induced metric is given by (\ref{eq:deltaq.def}).

Note that the boundary conditions formulated in this section should be supplemented with regularity of physically relevant, geometrical quantities such as the Ricci scalar of the induced metric on the brane, the tetrad components of the bulk Weyl tensor evaluated on the brane, etc.
This is because we are adopting the thin brane approximation and all we can and should trust is what is obtained within the validity of this approximation.
If, e.g., the Ricci scalar of the induced metric were singular then our approximation would be invalidated.
It is of course possible to regularize the singularity by introducing a finite thickness of the brane.
However, in this case the natural cutoff of the low-energy effective theory is the inverse of the thickness, and in general the regularized `would-be singularity' is not expected to be below the cutoff scale.
This simply means that we need a UV completion, e.g., the microphysical description of the brane, to describe the physics of the regularized `would-be singularity'.
Therefore, in general we have two options: (i) to specify a fundamental theory such as string theory as a UV completion and go on, or (ii) to concentrate on modes which are within the validity of the effective theory.
In the present approach we adopt the latter attitude, assuming the existence of a good UV completion but never using its properties.
This is the reason why we adopt the thin brane approximation and require the regularities.

In the next section, we shall apply these expressions to perturbations of the six-dimensional braneworld model explained in the introduction.

%%%%%%%%%%%%%%%%%%%%%%%%%%%%%%%%%%%%%%%%%%%%%%%%%%%%%%%%%%%%%%%%%%%%%%%%
\section{
Application to codimension-2 braneworld
}
\label{sec:appl}

Now we apply the formalism developed in the previous section to a concrete example.
We investigate perturbation of a six-dimensional braneworld model with warped flux compactification.
Two families of background solutions were introduced in section \ref{sec:intro}.
The first family is given by (\ref{eq:bg.gen}) together with (\ref{eq:bg.dS}), where $ q_{\mu\nu} $ is a four-dimensional Einstein space with the effective cosmological constant $ 3 $.
The second family is obtained as a limit of the first family and is specified by (\ref{eq:bg.gen}) with (\ref{eq:bg.flat}), where $ q_{\mu\nu} $ is an Einstein space with vanishing effective cosmological constant.
As explained in the introduction, these families of solutions have many interesting aspects as a model of braneworld.

For these backgrounds, the Killing vector is 
\begin{equation}
\varphi^{(0)}{}^M = \partial_\phi^M,
\end{equation}
corresponding to the angular coordinate $ \phi $.
We denote the period of the angular coordinate by $ \Delta\phi $, i.e., $ \phi \sim \phi + \Delta\phi $.
Since $ L^{(0)} $ is written as 
\begin{equation}
L^{(0)}
 = \sqrt{g^{(0)}_{\phi\phi}}
 = \sqrt{f(r)},
\end{equation}
the boundary condition (\ref{eq:L}) for the background says that the unperturbed position $ r_\mathrm b^{(0)} $ of the brane is a zero of the metric function $ f(r) $, as mentioned in the introduction.
The unit normal to the brane is 
\begin{equation}
n^{(0)}{}^M
 = \mp \sqrt f \partial_r^M,
\label{eq:n0.example}
\end{equation}
where the minus and plus signs are for the cases $ f'(r_\mathrm b^{(0)}) < 0 $ and $ f'(r_\mathrm b^{(0)}) > 0 $, respectively\footnote{
Note that we have chosen the direction of $n^M$ so that it directs from the brane to the bulk.
Thus, the sign convention here was chosen so that, when the bulk is bounded by two roots $ r_\pm $ ($ r_- < r_+ $) of $ f $, the minus sign in (\ref{eq:n0.example}) is for the larger root $ r_+ $ and the plus sign is for the smaller root $ r_- $.
We do not consider the degenerate case with $ f'(r_\mathrm b) = 0 $.
Actually, within the family of solutions considered here, there is no regime of parameters that gives a degenerate root of $ f $ and compact extra-dimensions.
}.
Thus, we obtain
\begin{equation}
\partial_\perp^{(0)} L^{(0)}
 = \mp\sqrt f \partial_r \sqrt f
 = \mp\frac{1}{2} f'(r).
\end{equation}
The corresponding boundary condition (\ref{eq:dL}) relates the parameters of the solution to the brane tension $ \sigma_\mathrm b $ as
\begin{equation}
\mp\frac{1}{2} f'(r_\mathrm b^{(0)})
 = \frac{2\pi-\delta_\mathrm b}{\Delta\phi}, \quad
\delta_\mathrm b = \frac{\sigma_\mathrm b}{M_6^4}.
\end{equation}
For appropriate values of brane tension, we can put one or two de Sitter or Minkowski branes at zeros of the function $ f(r) $ by imposing this boundary condition.
For more details of the properties of the solutions, see \cite{Mukohyama:2005yw}.

Linear perturbation around these backgrounds is described by $ \delta g_{MN} $, $ \delta A_M $ and $ \delta Z^M $.
(We have set $ \delta\varphi^M = 0 $ and $ \delta(\Delta\varphi) = 0 $.
See the third paragraph of section \ref{sec:formalism.perturb.bc}, especially the sentences just after equation (\ref{eq:deltae.def}).)
As already stated in the formulation, we still assume that the bulk spacetime has an axisymmetry and that the brane is at the center of the symmetry.
Hence, $ \delta Z^M $ can be written as 
\begin{equation}
\delta Z^M = \delta r_\mathrm b \partial_r^M.
\end{equation}
The boundary condition is given by (\ref{eq:bc.perturb}), where $ \delta\mcl_{0,1,2} $ and $ \delta\mca $ are calculated by the formulas (\ref{eq:deltaL0.simple}), (\ref{eq:deltaL1.simple}), (\ref{eq:deltaL2.simple}). and (\ref{eq:deltaA.simple}) as
\begin{align}
\begin{split}
\delta\mcl_0
 & = \left[
     \delta g_{\phi\phi} + f' \delta r_\mathrm b
     \right]_{r=r_\mathrm b^{(0)}}, \\
\delta\mcl_1
 & = \left[
     -\frac{1}{2} f^{3/2} f' \delta g_{rr}
     + \sqrt{f} \delta g_{\phi\phi}'
     + \left(\sqrt f f'\right)' \delta r_\mathrm b
     \right]_{r=r_\mathrm b^{(0)}}, \\
\delta\mcl_2
 & = \left[
     -f^{3/2} \left(\sqrt f f'\right)' \delta g_{rr}
     - \frac{1}{2} f f' \left(f \delta g_{rr}\right)'
     + \sqrt f \left(\sqrt f \delta g_{\phi\phi}'\right)'
     + \left(\sqrt f \left(\sqrt f f'\right)'\right)'
       \delta r_\mathrm b
     \right]_{r=r_\mathrm b^{(0)}}, \\
\delta\mca
 & = \left[\delta A_\phi + A' \delta r_\mathrm b\right]_{r=r_\mathrm b^{(0)}}.
\end{split}
\label{eq:bc.wfc}
\end{align}
Here, a prime denotes derivative with respect to $ r $.
The second-order derivative $ \delta g_{\phi\phi}'' $ can actually be eliminated by using the equation of motion.

%%%%%%%%%%%%%%%%%%%%%%%%%%%%%%%%%%%%%%%%%%%%%%%%%%%%%%%%%%%%%%%%%%%%%%%%
\subsection{
Scalar perturbations in the 6D braneworld and the boundary condition
}
\label{sec:appl.lowe}

The boundary condition given above is applicable to perturbations around both families of background solutions.
In the following, we shall restrict our consideration to the second class of background solutions, i.e., (\ref{eq:bg.gen}) and (\ref{eq:bg.flat}), and set $ q_{\mu\nu} $ to be the Minkowski metric.
Linear perturbations around background solutions in this sub-class were studied extensively in our previous work \cite{Yoshiguchi:2005nn}.
In particular, perturbative stability was shown and the Kaluza--Klein spectra were obtained.

In \cite{Yoshiguchi:2005nn}, the boundary condition for scalar-type perturbations was not derived from the first principle but obtained by assuming that a particular set of perturbation variables and their derivatives with respect to the coordinate $ r $ should be finite on the brane.
However, it is not {\it a priori} clear which variables should be finite on the brane.
Suppose we have two perturbation variables $ \Phi_1 $ and $ \Phi_2 $.
It is also possible to use another combination, say $ r^2 \Phi_1 $ and $ \Phi_2 + r \Phi_1' $.
Indeed, there are in principle infinite possibilities for the choice of perturbation variables.
The finiteness of which set of perturbation variables should we impose?
Even if we can somehow guess the answer to this question, there are in principle infinite possibilities for the choice of the coordinate with respect to which the derivatives of the perturbation variables are set to be finite on the brane.
Therefore, it is crucial to justify the boundary condition used in the analysis of \cite{Yoshiguchi:2005nn} from the first principle.
This is what we shall do in the following.
Moreover, in section \ref{sec:appl.stab} we shall analytically show the stability of the background in a particular limit called the football-shape limit.

Here we concentrate on scalar-type perturbations with the gauge choice adopted in \cite{Yoshiguchi:2005nn}:
\begin{align}
\begin{split}
g_{MN} \mathrm dx^M \mathrm dx^N
 & = (1+\Psi Y) r^2 \eta_{\mu\nu} \mathrm dx^\mu \mathrm dx^\nu
     + 2 h_{(\mathrm L)\phi} V_{(\mathrm L)\mu} \mathrm dx^\mu \mathrm d\phi
     + [1+(\Phi_1+\Phi_2) Y] \frac{\mathrm dr^2}{f}
     + [1-(\Phi_1+3\Phi_2) Y] f \mathrm d\phi^2, \\
A_M \mathrm dx^M
 & = a_r Y \mathrm dr
     + (A + a_\phi Y) \mathrm d\phi.
\end{split}
\end{align}
Here the metric components in the right-hand side have been already expanded by the harmonics of the Minkowski space summarized in Appendix~\ref{app:harmonics}, and integration in the $ k $ space is understood although it is not written explicitly.
For completeness, Appendix~\ref{app:gauge} includes detailed explanation of the gauge transformation leading to this gauge choice.
The coefficients $ \{\Phi_1,\Phi_2,\Psi,h_{(\mathrm L)\phi},a_r,a_\phi\} $ are functions of $ r $ and represent physical perturbations up to a residual gauge freedom $ C' $ explained in Appendix~\ref{app:gauge}.
We also expand $ \delta r_\mathrm b $ as
\begin{equation}
\delta r_\mathrm b
 = \int\mathrm dk \widetilde{\delta r_\mathrm b} Y.
\end{equation}
Correspondingly, by using (\ref{eq:bc.wfc}), we obtain the harmonic expansion of quantities $ \delta\mcl_{0,1,2} $ and $ \delta\mca $ relevant for the perturbed boundary condition (\ref{eq:bc.perturb}): 
\begin{equation}
\delta\mcl_{0,1,2}
 = \int\mathrm dk \widetilde{\delta\mcl}_{0,1,2} Y, \quad
\delta\mca
 = \int\mathrm dk \widetilde{\delta\mca} Y, 
\end{equation}
where 
\begin{align}
\begin{split}
\widetilde{\delta\mcl}_0
 & = \left[
     -f (\Phi_1 + 3 \Phi_2) + f' \widetilde{\delta r_\mathrm b}
     \right]_{r=r_\mathrm b^{(0)}}, \\
\widetilde{\delta\mcl}_1
 & = \left[
     -\frac{1}{2} \sqrt f f' (3\Phi_1 + 7\Phi_2)
     - f^{3/2} (\Phi_1' + 3\Phi_2')
     + \frac{f'^2+2f f''}{2 \sqrt f} \widetilde{\delta r_\mathrm b}
     \right]_{r=r_\mathrm b^{(0)}}, \\
\widetilde{\delta\mcl}_2
 & = \left[
     -(f'^2+2 f f'') (\Phi_1+2\Phi_2)
     - f f' (3\Phi_1'+8\Phi_2')
     - f^2 (\Phi_1''+3\Phi_2'')
     + (2 f' f''+f f''') \widetilde{\delta r_\mathrm b}
     \right]_{r=r_\mathrm b^{(0)}}, \\
\widetilde{\delta\mca}
 & = \left[a_\phi + A' \widetilde{\delta r_\mathrm b}\right]_{r=r_\mathrm b^{(0)}}.
\end{split}
\label{eq:bc.before_eom}
\end{align}
Thus, the perturbed boundary condition becomes
\begin{equation}
\widetilde{\delta\mcl}_{0,1,2} = \widetilde{\delta\mca} = 0.
\end{equation}

In the following we shall simplify the expression (\ref{eq:bc.before_eom}) and show that this boundary condition is actually equivalent to the naive boundary condition used in \cite{Yoshiguchi:2005nn}.

For this purpose, first, we shall use the field equations in the bulk.
As shown in \cite{Yoshiguchi:2005nn}, we can algebraically solve a part of Einstein and Maxwell equations to express four of six variables in terms of the other two:
\begin{equation}
\Psi = \Phi_2, \quad
h_{(\mathrm L)\phi} = C f, \quad
a_r = -\frac{A'}{f} h_{(\mathrm L)\phi} = -C A', \quad
a_\phi = \frac{1}{A'}
         \left[f' \Phi_2 + \frac{1}{2r^2} (f r^2 \Phi_1)'\right],
\label{eq:sol.scalar.constraint}
\end{equation}
where $ C $ is a constant corresponding to the residual gauge represented by $ C' $ in Appendix~\ref{app:gauge}.
The remaining field equations reduce to
\begin{align}
\begin{split}
& \Phi_2'' + \frac{4}{r} \Phi_2' + \frac{m^2}{2 r^2 f} (\Phi_1+2\Phi_2) = 0, \\
& \Phi_1'' + 2 \left(\frac{f'}{f}+\frac{5}{r}\right) \Phi_1'
  - \frac{4\Lambda_6}{f} (\Phi_1+\Phi_2) + \frac{m^2}{r^2 f} \Phi_1 = 0,
\label{eq:eom.scalar}
\end{split}
\end{align}
where $ m^2 = -\eta^{\mu\nu} k_\mu k_\nu $.
By using these equations of motion, $ \widetilde{\delta\mcl}_2 $ in (\ref{eq:bc.before_eom}) is reduced to
\begin{multline}
\widetilde{\delta\mcl}_2
 = \biggl[
   -\left(f'^2 + f\left(4\Lambda_6-\frac{5m^2}{2r^2}+2f''\right)\right) \Phi_1
   - \left(2f'^2 + f\left(4\Lambda_6-\frac{3m^2}{r^2}+4f''\right)\right) \Phi_2
   \biggr. \\
   \biggl.
   - f\left(f'-\frac{10f}{r}\right) \Phi_1'
   - 4f\left(2f'-\frac{3f}{r}\right) \Phi_2'
   + (2 f' f''+f f''') \widetilde{\delta r_\mathrm b}
   \biggr]_{r=r_\mathrm b^{(0)}}.
\label{eq:bc.after_eom}
\end{multline}

Second, as explained in the end of section \ref{sec:formalism}, we require that geometrical quantities be regular on the brane.
In particular, we require that linear perturbations of $ R $, $ R^{MNPQ} R_{MNPQ} $, $ \nabla^M R \nabla_M R $, and $ \nabla^R R^{MNPQ} \nabla_R R_{MNPQ} $ to be regular on the boundaries, where $ f $ vanishes.
By using computer algebra packages such as {\tt GRTensor} \cite{GRTensor}, we can express these geometrical quantities in terms of $ \Phi_{1,2} $ and their derivatives.
After eliminating second and higher order derivatives by (\ref{eq:eom.scalar}), they are reduced to linear combinations of just $ \Phi_{1,2} $ and $ \Phi_{1,2}' $, with coefficients being functions of $ r $.
At this point, however, the matrix made of the coefficients does not have a regular inverse in the limit $ f \to 0 $.
After some algebra, it is found that the linear transformation from $ \mcr_{1,2,3,4} $ to linear perturbations of $ R $, $ R^{MNPQ} R_{MNPQ} $, $ \nabla^M R \nabla_M R $, and $ \nabla^R R^{MNPQ} \nabla_R R_{MNPQ} $ is regular and has a regular inverse in the limit $ r \to r_\mathrm b^{(0)} $, where $ \mcr_{1,2,3,4} $ are defined as $ \mcr_1 \equiv \Phi_2 $, $ \mcr_2 \equiv f \Phi_1 $, $ \mcr_3 \equiv f' \Phi_2'-(m^2/2r^2) \Phi_1 $, and $ \mcr_4 \equiv (f\Phi_1)' $.
Therefore, $ \mcr_{1,2,3,4} $ should be regular on the boundary.
Since $ \mcr_{1,2,3,4} $ multiplied by positive powers of $ f $ vanish on the boundary, let us express $ \widetilde{\delta\mcl}_{0,1,2} $ in terms of $ \mcr_{1,2,3,4} $ instead of $ \Phi_{1,2} $ and $ \Phi_{1,2}' $ in order to identify terms remaining on the boundary.
The result is 
\begin{align}
\begin{split}
\widetilde{\delta\mcl}_0
 & = \left[
     -\mcr_2 - 3f \mcr_1 + f' \widetilde{\delta r_\mathrm b}
     \right]_{r=r_\mathrm b^{(0)}}, \\
\widetilde{\delta\mcl}_1
 & = \left[
     - \left(\frac{3m^2 \sqrt f}{2r^2f'}+\frac{f'}{2\sqrt f}\right) \mcr_2
     - \frac{7}{2} \sqrt f f' \mcr_1
     - \sqrt f \mcr_4
     - \frac{3f^{3/2}}{f'} \mcr_3
     + \left(\frac{f'^2}{2\sqrt f}+\sqrt f f''\right)
       \widetilde{\delta r_\mathrm b}
     \right]_{r=r_\mathrm b^{(0)}}, \\
\widetilde{\delta\mcl}_2
 & = \biggl[
     - \left(4\Lambda_6+\frac{3m^2}{2r^2}+\frac{10f'}{r}
             +2f''-\frac{6m^2 f}{r^3 f'}\right) \mcr_2
     - \left(2f'^2+ f \left(4\Lambda_6-\frac{3m^2}{r^2}+4f''\right)\right) \mcr_1
     \biggr. \\
 &   \qquad
     \biggl.
     - \left(f'-\frac{10f}{r}\right) \mcr_4
     - 4 f \left(2-\frac{3f}{rf'}\right) \mcr_3
     + (2 f' f''+f f''') \widetilde{\delta r_\mathrm b}
     \biggr]_{r=r_\mathrm b^{(0)}}.
\end{split}
\label{eq:bc.after_eom_regular}
\end{align}
Since $ \mcr_{1,2,3,4} $ are regular on the brane, all terms in the above expressions are manifestly regular.
Terms with positive powers of $ f $ all vanish on the brane and, thus, we obtain the following expressions after recovering $ \Phi_{1,2} $ and $ \Phi_{1,2}' $.
\begin{align}
\begin{split}
\widetilde{\delta\mcl}_0
 & = \left[
     -f \Phi_1 + f' \widetilde{\delta r_\mathrm b}
     \right]_{r=r_\mathrm b^{(0)}}, \\
\widetilde{\delta\mcl}_1
 & = \left[
     -\frac{\sqrt f f'}{2} \Phi_1
     + \frac{f'^2}{2\sqrt f} \widetilde{\delta r_\mathrm b}
     \right]_{r=r_\mathrm b^{(0)}}, \\
\widetilde{\delta\mcl}_2
 & = \left[
     -\left(4\Lambda_6+\frac{3m^2}{2r^2}+\frac{10f'}{r}+2f''\right) f \Phi_1
     - 2 f'^2 \Phi_2 - f' (f \Phi_1)'
     + 2 f' f'' \widetilde{\delta r_\mathrm b}
     \right]_{r=r_\mathrm b^{(0)}}.
\end{split}
\end{align}
Actually, the four boundary conditions $ \widetilde{\delta\mcl}_{0,1,2} = \widetilde{\delta\mca} = 0 $ are not independent but only three of them are independent since $ a_\phi $ is now written in terms of $ \Phi_{1,2} $ as in (\ref{eq:sol.scalar.constraint}).
Hence the independent conditions are
\begin{equation}
\left.f \Phi_1\right|_{r=r_\mathrm b^{(0)}}
 = \left[2 f' \Phi_2 + \left(f \Phi_1\right)'\right]_{r=r_\mathrm b^{(0)}}
 = 0,
\label{eq:bc.scalar}
\end{equation}
and 
\begin{equation}
\widetilde{\delta r_\mathrm b} = 0.
\end{equation}
The last equation means that the brane position in this gauge is not changed by the perturbation.
Note, however, that we would not be able to know this fact unless we used the doubly covariant formalism developed in this paper.
With the manifest replacements $ r_\mathrm b^{(0)} \to r_\pm^{(0)} $ and $ \widetilde{\delta r_\mathrm b} \to \widetilde{\delta r_\pm} $, the boundary condition (\ref{eq:bc.scalar}) is exactly what we obtained in \cite{Yoshiguchi:2005nn} by naively setting coefficients of $ f^{-1} $ to be zero.
On the other hand, in the present paper we have reached the same boundary condition from the first principle.
Therefore, we have justified the treatment in \cite{Yoshiguchi:2005nn}.

For vector and tensor perturbations, the perturbed boundary condition (\ref{eq:bc.perturb}) does not give any requirement since the quantities $ \mcl_{0,1,2} $ and $ \mca $ are scalars.
Nonetheless, as shown in \cite{Yoshiguchi:2005nn}, regularity of geometrical quantities on the brane gives sufficient boundary conditions.

%%%%%%%%%%%%%%%%%%%%%%%%%%%%%%%%%%%%%%%%%%%%%%%%%%%%%%%%%%%%%%%%%%%%%%%%
\subsection{
Stability of the football-shape limit
}
\label{sec:appl.stab}

Next we consider a limit in which the shape of the extra dimension becomes locally a round $ 2 $-sphere.
Taking the $ \alpha \equiv r_-/r_+ \to 1 $ limit gives
\begin{equation}
g^{(0)}{}_{MN} \mathrm dx^M \mathrm dx^N
 = r_0^2 \eta_{\mu\nu} \mathrm dx^\mu \mathrm dx^\nu
 + R^2 \Omega_{ab} \mathrm dx^a \mathrm dx^b,
\label{eq:bg.sph}
\end{equation}
where the constant $ r_0 $ represents the position of the two branes and $ R $ is the radius of the sphere related to the bulk cosmological constant as $ R = 1/\sqrt{2\Lambda_6} $.
$ \Omega_{ab} $ is the metric of round sphere and $ \varphi $ parameterizes its azimuthal angle whose period is $ 2\pi-\delta $, i.e., the sphere has a deficit angle $ \delta = \sigma/M_6^4 $, where $ \sigma $ is the common tension of the branes.
Although the coordinate distance $ r_+ - r_- $ between the branes vanishes and the bulk geometry seems to disappear, the proper distance remains finite.
For the detail of this limit, see Appendix~\ref{app:sphere}.
In the following we normalize the warp factor at the position of the branes and the radius of the sphere so that $ r_0 = R = 1 $.
The geometry of the extra dimensions has a coordinate representation
\begin{align}
\begin{split}
\Omega_{ab} \mathrm dx^a \mathrm dx^b
 & = \frac{\mathrm dw^2}{\bar f(w)} + \bar f(w) \mathrm d\varphi^2, \\
A_M \mathrm dx^M
 & = \bar A(w) \mathrm d\varphi,
\end{split}
\end{align}
in which the metric function and the gauge potential are specified as
\begin{equation}
\bar f = 1-w^2, \quad \bar A = -w.
\end{equation}
The background quantities are given as
\begin{equation}
\varphi^{(0)}{}^M
 = \partial_\varphi^M, \quad
L^{(0)}
 = \sqrt{1-w^2}, \quad
n^{(0)}{}^M
 = \sqrt{1-w^2} \partial_w^M, \quad
q^{(0)}{}_{\mu\nu}
 = \eta_{\mu\nu}, \quad
w_\pm^{(0)} = \pm 1.
\end{equation}
Perturbations are defined on this background.
Schematically, all the formalism developed above can be directly applied to this limit together with the following replacements 
\begin{equation}
(r,\phi) \to (w,\varphi), \quad
f \to \bar f, \quad
A \to \bar A.
\end{equation}

In the following, we will show stability of all the types of perturbation for completeness.
However, the boundary conditions obtained above are relevant for the scalar perturbations only.
Thus if the reader is particularly interested in direct applications, it is recommended to skip to section \ref{sec:appl.stab.scalar}.
As will be shown there, the boundary conditions for the scalar perturbations at the branes (\ref{eq:bc.scalar}) are further reduced to 
\begin{equation}
\left[\Phi_1+2 \Phi_2\right]_{w=\pm1} = 0.
\label{eq:bc.scalar_pre}
\end{equation}

%%%%%%%%%%%%%%%%%%%%%%%%%%%%%%%%%%%%%%%%%%%%%%%%%%%%%%%%%%%%%%%%%%%%%%%%
\subsubsection{
Tensor perturbation
}

The configuration with tensor perturbation is given by 
\begin{align}
\begin{split}
g_{MN} \mathrm dx^M \mathrm dx^N
 & = (\eta_{\mu\nu} + h_{(\mathrm T)} T_{(\mathrm T)\mu\nu})
     \mathrm dx^\mu \mathrm dx^\nu
     + \frac{\mathrm dw^2}{1-w^2} + (1-w^2) \mathrm d\varphi^2, \\
A_M \mathrm dx^M
 & = -w \mathrm d\varphi,
\end{split}
\end{align}
where $ h_{(\mathrm T)} $ denotes physical perturbation.
Tensor perturbation is gauge-invariant.

The Einstein equations are reduced to
\begin{equation}
[(1-w^2) h_{(\mathrm T)}']' + \mu^2 h_{(\mathrm T)} = 0,
\end{equation}
where the prime denotes derivative over $ w $ and the (dimensionless) Kaluza--Klein mass squared $ \mu^2 $ is defined by $ \mu^2 = -\eta^{\mu\nu} k_\mu k_\nu $.
The solution is obtained as
\begin{equation}
h_{(\mathrm T)} = c P_\nu(w) + d Q_\nu(w),
\end{equation}
where $ P_\nu $ and $ Q_\nu $ are the Legendre functions of the first and second kind, respectively, and $ c $ and $ d $ are arbitrary constants.
Their order $ \nu $ is expressed by above defined $ \mu^2 $ as
\begin{equation}
\nu = \frac{-1+\sqrt{4\mu^2+1}}{2}.
\end{equation}
The condition imposed on the tensor perturbation is regularity on the boundaries $ w = \pm 1 $.
By requiring the Ricci scalar of the induced metric and vielbein components of the Weyl tensor to be regular there, we find that $ h_{(\mathrm T)} $ and $ h_{(\mathrm T)}' $ should be both regular on the boundaries.
The set of independent regular quantities used here is listed in Appendix~\ref{app:regular}.
The regularity of $ h_{(\mathrm T)} $ at $ w = 1 $ requires $ d = 0 $, and that at $ w = -1 $ does $ P_\nu $ itself to be non-singular there, which is realized when $ \nu $ is non-negative integer: $ \nu = 0,1,2,\ldots $.
The mass spectrum of the tensor perturbation is then determined as
\begin{equation}
\mu^2 = \nu (\nu+1) \qquad (\nu = 0,1,2,\ldots).
\end{equation}
The spectrum of the first few KK modes is listed in Table~\ref{tab:KK.tensor}.
Note that for simplicity we have been setting the radius of the extra dimensions as $ R = 1/\sqrt{2\Lambda_6} = 1 $, however, in general cases the mass spectrum is given by $ 2\Lambda_6 \mu^2 $ and this scaling applies also to the vector and scalar perturbations discussed in what follows.
\begin{center}
\begin{table}[ht]
\caption{The KK spectrum of the tensor perturbation.}
\label{tab:KK.tensor}
\begin{ruledtabular}
\begin{tabular}{c||c|c|c|c|c|c|c}
Level $ \nu $
 & $ 0 $ 
 & $ 1 $
 & $ 2 $
 & $ 3 $
 & $ 4 $
 & $ 5 $
 & $ \ldots $ \\
\hline
Mass $ \mu^2 $
 & $ 0 $
 & $ 2 $
 & $ 6 $
 & $ 12 $
 & $ 20 $
 & $ 30 $
 & $ \ldots $
\end{tabular}
\end{ruledtabular}
\end{table}
\end{center}

There is a zero mode describing the four-dimensional gravitational wave and other modes have positive $ \mu^2 $ hence we conclude that the spacetime is stable against linear tensor perturbation.

%%%%%%%%%%%%%%%%%%%%%%%%%%%%%%%%%%%%%%%%%%%%%%%%%%%%%%%%%%%%%%%%%%%%%%%%
\subsubsection{
Vector perturbation
}
\label{sec:appl.stab.vector}

The configuration with vector perturbation is given by
\begin{align}
\begin{split}
g_{MN} \mathrm dx^M \mathrm dx^N
 & = \eta_{\mu\nu} \mathrm dx^\mu \mathrm dx^\nu
     + 2 h_{(\mathrm T)a} V_{(\mathrm T)\mu} \mathrm dx^\mu \mathrm dx^a
     + \frac{\mathrm dw^2}{1-w^2} + (1-w^2) \mathrm d\varphi^2 \\
A_M \mathrm dx^M
 & = a_{(\mathrm T)} V_{(\mathrm T)\mu} \mathrm dx^\mu - w \mathrm d\varphi,
\end{split}
\end{align}
where $ \{h_{(\mathrm T)w},h_{(\mathrm T)\varphi},a_{(\mathrm T)}\} $ denote physical perturbations.
Gauge fixing is summarized in Appendix~\ref{app:gauge}.

First, the $ \mu\nu $ components of the Einstein equation is solved to give
\begin{equation}
h_{(\mathrm T)w} = \frac{D}{1-w^2},
\end{equation}
where $ D $ is an arbitrary constant when $ \mu^2 = -\eta^{\mu\nu} k_\mu k_\nu = 0 $, or $ D = 0 $ when $ \mu^2 \neq 0 $.
However, if $ D $ has a non-zero value, divergence of the metric components on the boundaries cannot be removed since no gauge freedom is remaining.
Hence we conclude that $ D = 0 $.
Remaining Einstein's equation and Maxwell equation reduce to
\begin{align}
\begin{split}
& (1-w^2) (h_{(\mathrm T)\varphi}''
  - 2 a_{(\mathrm T)}') + \mu^2 h_{(\mathrm T)\varphi}
  = 0, \\
& [(1-w^2) a_{(\mathrm T)}']'
  + h_{(\mathrm T)\varphi}' + \mu^2 a_{(\mathrm T)}
  = 0.
\end{split}
\label{eq:eom.vector.sph.pre}
\end{align}
By requiring the vielbein components of the gauge field strength and those of the Weyl tensor to be regular on the boundaries $ w = \pm 1 $, we find that $ a_{(\mathrm T)} $, $ \sqrt{1-w^2} a_{(\mathrm T)}' $, $ h_{(\mathrm T)\varphi}/(1-w^2) $, and $ h_{(\mathrm T)\varphi}' $ should be regular on the boundaries.
The set of independent regular quantities used here is listed in Appendix~\ref{app:regular}.

In the case of $ \mu^2 = 0 $, $ h_{(\mathrm T)\varphi}' $ is solved as
\begin{equation}
h_{(\mathrm T)\varphi}' = c P_1(w) + d Q_1(w),
\end{equation}
where $ c $ and $ d $ are constants, and $ P_1 $ and $ Q_1 $ are, respectively, Legendre polynomials of the first and second kind of order $ 1 $.
Requiring the regularity of $ h_{(\mathrm T)\varphi}/(1-w^2) $ and $ h_{(\mathrm T)\varphi}' $ at $ w = \pm 1 $, we obtain the zero mode of vector perturbation,
\begin{equation}
h_{(\mathrm T)\varphi} = c_1 (1-w^2), \quad
a_{(\mathrm T)} = - c_1 w + c_2,
\end{equation}
where $ c_i $'s are arbitrary constants corresponding to two degrees of freedom for the zero mode of vector perturbation.

Next we consider Kaluza--Klein modes.
When $ \mu^2 \neq 0 $, introducing following variables makes the analysis easier:
\begin{equation}
h_{(\mathrm T)\varphi}
 \equiv \frac{1}{\mu^2} (1-w^2) \Phi_{(\mathrm V)2}', \quad
a_{(\mathrm T)}
 \equiv \Phi_{(\mathrm V)1}.
\end{equation}
Then the field equations (\ref{eq:eom.vector.sph.pre}) become
\begin{align}
\begin{split}
& [(1-w^2) \Phi_{(\mathrm V)2}']'
 + \mu^2 (\Phi_{(\mathrm V)2} - 2 \Phi_{(\mathrm V)1}) = 0, \\
& [(1-w^2) \Phi_{(\mathrm V)1}']'
 + (\mu^2+2) \Phi_{(\mathrm V)1} - \Phi_{(\mathrm V)2} = 0,
\end{split}
\label{eq:eom.vector.sph}
\end{align}
where $ \Phi_{(\mathrm V)2}''' $ was removed by the equations of motion.
Eliminating $ \Phi_{(\mathrm V)1} $ from these equations gives a single fourth-order differential equation for $ \Phi_{(\mathrm V)2} $
\begin{equation}
[(1-w^2) E_\pm']' + \lambda_\pm E_\pm = 0,
\end{equation}
where
\begin{align}
\begin{split}
E_\pm
 & = [(1-w^2) \Phi_{(\mathrm V)2}']' + \lambda_\mp \Phi_{(\mathrm V)2}, \\
\lambda_\pm
 & = \mu^2+1 \pm \sqrt{2\mu^2+1}.
\end{split}
\end{align}
The case of $ \mu^2 = -1/2 $ causes the eigenvalues $ \lambda_\pm $ degenerate and requires another treatment.
If $ \mu^2 \neq -1/2 $, the solution for $ \Phi_{(\mathrm V)2} $ is obtained as a linear combination of the Legendre functions and, from (\ref{eq:eom.vector.sph}), $ \Phi_{(\mathrm V)1} $ is readily solved.
The general solutions are
\begin{align}
\begin{split}
\Phi_{(\mathrm V)2}
 & = c_+ P_{\nu_+} + c_- P_{\nu_-} + d_+ Q_{\nu_+} + d_- Q_{\nu_-}, \\
\Phi_{(\mathrm V)1}
 & = -\frac{1}{2\mu^2} \left[
     c_+ (\lambda_+-\mu^2) P_{\nu_+}
     + c_- (\lambda_--\mu^2) P_{\nu_-}
     + d_+ (\lambda_+-\mu^2) Q_{\nu_+}
     + d_- (\lambda_--\mu^2) Q_{\nu_-}
     \right],
\end{split}
\end{align}
where $ c_\pm, d_\pm $ are arbitrary constants and the values of order $ \nu_\pm $ are determined by
\begin{equation}
\nu_\pm (\nu_\pm+1) = \lambda_\pm.
\end{equation}
Now $ \Phi_{(\mathrm V)2} $, $ \Phi_{(\mathrm V)2}' $, $ \Phi_{(\mathrm V)1} $, and $ \sqrt{1-w^2} \Phi_{(\mathrm V)1}' $ are independent regular variables on the boundaries.
At $ w = 1 $, $ \Phi_{(\mathrm V)2} $ is regular only if $ d_+ + d_- = 0 $ whereas the regularity of $ \Phi_{(\mathrm V)1} $ requires $ d_+ (\lambda_+-\mu^2) + d_- (\lambda_--\mu^2) = 0 $.
Since $ \lambda_+ \neq \lambda_- $ for $ \mu^2 \neq -1/2 $, we obtain $ d_\pm = 0 $.
At $ w = -1 $, the regularity of the variables requires
\begin{align}
\begin{split}
0 & = c_+ \sin\nu_+\pi + c_- \sin\nu_-\pi, \\
0 & = c_+ \lambda_+ \sin\nu_+\pi + c_- \lambda_- \sin\nu_-\pi.
\end{split}
\end{align}
With $ \lambda_+ \neq \lambda_- $, this implies that non-trivial solutions can exist if $ \nu_+ \in \mathbb Z $ or $ \nu_- \in \mathbb Z $.
We have to only consider non-negative $ \nu_\pm $'s so that these conditions are explicitly written down as
\begin{equation}
\nu_\pm
 = \frac{-1+\sqrt{4\mu^2+5 \pm 4\sqrt{2\mu^2+1}}}{2} = 0,1,2,\ldots.
\end{equation}
This condition gives the KK mass spectrum for the vector perturbation.
The first few of them is listed in Table~\ref{tab:KK.vector}.
\begin{center}
\begin{table}[ht]
\caption{The KK spectrum of the vector perturbation.}
\label{tab:KK.vector}
\begin{ruledtabular}
\begin{tabular}{c||c|c|c|c|c|c|c}
Level
 & $ \nu_- = 0, \nu_+ = 1 $
 & $ \nu_+ = 2 $
 & $ \nu_- = 1 $
 & $ \nu_+ = 3 $
 & $ \nu_- = 2 $
 & $ \nu_+ = 4 $
% & $ \nu_- = 3 $
 & $ \ldots $ \\
\hline
Mass $ \mu^2 $
 & $ 0 $
 & $ 2(3-\sqrt{3}) $
 & $ 4 $
 & $ 2(6-\sqrt{6}) $
 & $ 2(3+\sqrt{3}) $
 & $ 2(10-\sqrt{10}) $
% & $ 2(6+\sqrt{6}) $
 & $ \ldots $
\end{tabular}
\end{ruledtabular}
\end{table}
\end{center}
As mentioned earlier, the zero mode of vector perturbation has two degrees of freedom, thus there are $ \nu_+ = 1 $ and $ \nu_- = 0 $ at $ \mu^2 = 0 $.

In the special case of $ \mu^2 = -1/2 $ the eigenvalues $ \lambda_\pm $ degenerate and the order is given a value $ \nu = (\sqrt{3}-1)/2 $.
The general solutions are given by
\begin{align}
\begin{split}
\Phi_{(\mathrm V)2}
 & = \left[
     c P_\nu + c' \partial_\nu P_\nu
     + d Q_\nu + d' \partial_\nu Q_\nu
     \right]_{\nu=(\sqrt 3-1)/2}, \\
\Phi_{(\mathrm V)1}
 & = \left[
     c P_\nu + c' (\sqrt 3+\partial_\nu) P_\nu
     + d Q_\nu + d' (\sqrt 3+\partial_\nu) Q_\nu
     \right]_{\nu=(\sqrt 3-1)/2},
\end{split}
\end{align}
where $ c,c',d,d' $ are constants.
At $ w = 1 $, they are regular only if $ d = d' = 0 $.
Then, to remove divergence at $ w = -1 $, we also obtain $ c = c' = 0 $.
Therefore $ \mu^2 = -1/2 $ mode does not exist.

All the KK modes have positive $ \mu^2 $ hence we conclude the spacetime is stable against linear vector perturbation.

%%%%%%%%%%%%%%%%%%%%%%%%%%%%%%%%%%%%%%%%%%%%%%%%%%%%%%%%%%%%%%%%%%%%%%%%
\subsubsection{
Scalar perturbation
}
\label{sec:appl.stab.scalar}

Finally we investigate scalar perturbations.
Now it is required to consider boundary conditions derived by our formalism.
This in contrast with the other types of perturbations, for which the regularity of geometrical quantities was sufficient.

The configuration with scalar perturbation is given by
\begin{align}
\begin{split}
g_{MN} \mathrm dx^M \mathrm dx^N
 & = (1+\Psi Y) \eta_{\mu\nu} \mathrm dx^\mu \mathrm dx^\nu
     + 2 h_{(\mathrm L)\varphi} V_{(\mathrm L)\mu} \mathrm dx^\mu \mathrm d\varphi
     + [1+(\Phi_1+\Phi_2) Y] \frac{\mathrm dw^2}{1-w^2} \\
 & \qquad + [1-(\Phi_1+3\Phi_2) Y] (1-w^2) \mathrm d\varphi^2 \\
A_M \mathrm dx^M
 & = a_w Y \mathrm dw
     + (-w + a_\varphi Y) \mathrm d\varphi
\end{split}
\end{align}
where we have already chosen an analogue of the longitudinal gauge so that $ \{\Phi_1,\Phi_2,\Psi,h_{(\mathrm L)\varphi},a_w,a_\varphi\} $ denote physical perturbation.
Note that there remains a residual gauge freedom in $ h_{(\mathrm L)\varphi} $.
Gauge fixing is summarized in Appendix~\ref{app:gauge}.
From Einstein's equations, four of six variables are algebraically solved as
\begin{equation}
\Psi = \Phi_2, \quad
h_{(\mathrm L)\varphi} = C (1-w^2), \quad
a_w = -\frac{1}{2} h_{(\mathrm L)\varphi}'' = C, \quad
a_\varphi = w (\Phi_1+2\Phi_2) - \frac{1}{2} (1-w^2) \Phi_1',
\end{equation}
where $ C $ is a constant corresponding to the residual gauge, which is represented by $ C' $ in Appendix~\ref{app:gauge}.
Then the remaining field equations reduce to
\begin{align}
\begin{split}
& (1-w^2) \Phi_2'' + \frac{\mu^2}{2} (\Phi_1+2\Phi_2) = 0, \\
& (1-w^2) \Phi_1'' - 4w \Phi_1' - (2-\mu^2) \Phi_1 - 2 \Phi_2 = 0,
\end{split}
\label{eq:eom.scalar.sph}
\end{align}
where $ \mu^2 = -\eta^{\mu\nu} k_\mu k_\nu $.
By requiring the Ricci scalar, Kretschmann scalar, vielbein components of the Weyl tensor to be regular on the boundaries $ w = \pm 1 $, we find that $ \Phi_2 $, $ \Phi_2' $, $ \Phi_1 $, and $ \sqrt{1-w^2} \Phi_1' - (2w/\sqrt{1-w^2}) (\Phi_1+2\Phi_2) $ should be regular on the boundaries.
Multiplying the last one by $ \sqrt{1-w^2} $ and given that both $ \Phi_2 $ and $ \Phi_1 $ is regular at the boundaries, we deduce that $ (1-w^2) \Phi_1' $ is also regular.
The set of independent regular quantities used here is listed in Appendix~\ref{app:regular}.

Following analysis needs the aid of the boundary condition (\ref{eq:bc.wfc}).
By the knowledge of the regular variables, we can drop terms consist of regular combination multiplied by positive powers of $ 1-w^2 $.
Second order derivative of the metric component appearing in $ \widetilde{\delta\mcl}_2 $ is removed by the equation of motion (\ref{eq:eom.scalar.sph}).
After such reductions, we obtain the boundary conditions as shown in (\ref{eq:bc.scalar_pre}),
\begin{equation}
\left[\Phi_1 + 2\Phi_2\right]_{w=\pm1} = 0,
\label{eq:bc.scalar.sph}
\end{equation}
and the regularity at $ w = \pm 1 $ yields $ a_\varphi|_{w=\pm1} = 0 $.

Now we show the absence of zero mode.
When $ \mu^2 = 0 $, the bulk solution for (\ref{eq:eom.scalar.sph}) is
\begin{align}
\begin{split}
\Phi_2
 & = c_1 w + c_2, \\
\Phi_1
 & = \frac{c_1 w^3 + 3 c_2 w^2 + c_3 w + c_4}{3 (1-w^2)},
\end{split}
\end{align}
where $ c_i $'s are arbitrary constants.
This system has in fact no solution other than $ \Phi_1 = \Phi_2 = 0 $ when the regularity at the boundaries and the boundary condition (\ref{eq:bc.scalar.sph}) are imposed.

Next we consider Kaluza--Klein modes.
Eliminating $ \Phi_1 $ from the field equations (\ref{eq:eom.scalar.sph}) gives a single forth-order differential equation for $ \Phi_2 $
\begin{equation}
[(1-w^2) E_\pm']' + \lambda_\mp E_\pm = 0,
\end{equation}
where
\begin{align}
\begin{split}
E_\pm
 & = [(1-w^2) \Phi_2']' + \lambda_\pm \Phi_2, \\
\lambda_\pm
 & = \mu^2+1 \pm \sqrt{3\mu^2+1}.
\end{split}
\end{align}
The case of $ \mu^2 = -1/3 $ causes the eigenvalues $ \lambda_\pm $ degenerate and requires another treatment.
If $ \mu^2 \neq -1/3 $, the solution for $ \Phi_2 $ is obtained as a linear combination of the Legendre functions, and, from (\ref{eq:eom.scalar.sph}), $ \Phi_1 $ is readily solved.
The general solutions are
\begin{align}
\begin{split}
\Phi_2
 & = c_+ P_{\nu_+} + c_- P_{\nu_-} + d_+ Q_{\nu_+} + d_- Q_{\nu_-}, \\
\Phi_1
 & = \frac{2}{\mu^2}
     \bigl[
     c_+ (\lambda_+-\mu^2-2w \partial_w) P_{\nu_+}
     + c_- (\lambda_--\mu^2-2w \partial_w) P_{\nu_-}
     \bigr. \\
 &   \qquad\qquad
     \bigl.
     + d_+ (\lambda_+-\mu^2-2w \partial_w) Q_{\nu_+}
     + d_- (\lambda_--\mu^2-2w \partial_w) Q_{\nu_-}
     \bigr],
\end{split}
\label{eq:sol.scalar.sph}
\end{align}
where $ c_\pm,d_\pm $ are arbitrary constants and the values of order $ \nu_\pm $ are determined by
\begin{equation}
\nu_\pm (\nu_\pm+1) = \lambda_\pm.
\end{equation}
Again we insist that, for the scalar perturbation, it is not sufficient to impose regularity on the variables but the boundary condition is needed.
The boundary condition for $ \Phi_1 $ and $ \Phi_2 $ is expanded around $ w = 1 $ as
\begin{align}
0
 & = [\Phi_1 + 2\Phi_2]_{w=1} \nonumber\\
 & \simeq \frac{2}{\mu^2} (d_+ + d_-) (1-w)^{-1}
   - \frac{1}{\mu^2} [d_+ (\lambda_+-1) + d_- (\lambda_--1)] + \mco(1-w).
\end{align}
Since $ \lambda_+ \neq \lambda_- $, this means $ d_+ = d_- = 0 $.
Next we expand it around $ w = -1 $, then
\begin{align}
0
 & = [\Phi_1 + 2\Phi_2]_{w=-1} \nonumber\\
 & \simeq \frac{4}{\pi \mu^2} (c_+ \sin\nu_+\pi + c_- \sin\nu_-\pi) (1+w)^{-1}
   + \frac{2}{\pi \mu^2}
     [c_+ (\lambda_+-1) \sin\nu_+\pi + c_- (\lambda_--1) \sin\nu_-\pi]
   + \mco(1+w).
\end{align}
This, again with $ \lambda_+ \neq \lambda_- $, shows that non-trivial solutions can exist if $ \nu_+ \in \mathbb Z $ or $ \nu_- \in \mathbb Z $.
We have to only consider non-negative $ \nu_\pm $'s so that these conditions are explicitly written down as
\begin{equation}
\nu_\pm
 = \frac{-1+\sqrt{4\mu^2+5 \pm 4\sqrt{3\mu^2+1}}}{2} = 0,1,2,\ldots.
\end{equation}
Substituting (\ref{eq:sol.scalar.sph}) into the expression for $ a_\varphi $ with a non-negative integer $ \nu = \nu_\pm $, we can explicitly examine $ a_\varphi|_{w=\pm1} = 0 $.
Finally, the above condition for $ \nu $ gives the KK mass spectrum for scalar perturbation, which is listed in Table~\ref{tab:KK.scalar}.
\begin{center}
\begin{table}[ht]
\caption{The KK spectrum of the scalar perturbation.}
\label{tab:KK.scalar}
\begin{ruledtabular}
\begin{tabular}{c||c|c|c|c|c|c|c}
Level
 & $ \nu_- = 0 $ 
 & $ \nu_+ = 2 $
 & $ \nu_- = 1 $
 & $ \nu_+ = 3 $
 & $ \nu_- = 2 $
 & $ \nu_+ = 4 $
 & $ \ldots $ \\
\hline
Mass $ \mu^2 $
 & $ 1 $
 & $ (13-\sqrt{73})/2 $
 & $ 5 $
 & $ (25-\sqrt{145})/2 $
 & $ (13+\sqrt{73})/2 $
 & $ (41-\sqrt{241})/2 $
 & $ \ldots $
\end{tabular}
\end{ruledtabular}
\end{table}
\end{center}

In the case of $ \mu^2 = -1/3 $, the eigenvalues $ \lambda_\pm $ degenerate and the order is given a value $ \nu = (\sqrt{33}-3)/6 $.
The general solutions are given by
\begin{align}
\begin{split}
\Phi_2
 & = \left[c P_\nu + c' \partial_\nu P_\nu
           + d Q_\nu + d' \partial_\nu Q_\nu\right]_{\nu=\frac{\sqrt{33}-3}{6}}, \\
\Phi_1
 & = -2
     \Bigl[
     3 c (1-2w\partial_w) P_\nu
     + c' \left(\sqrt{33}+3(1-2w\partial_w)\partial_\nu\right) P_\nu
     \Bigr. \\
 &   \qquad\qquad
     \Bigl.
     + 3 d (1-2w\partial_w) Q_\nu
     + d' \left(\sqrt{33}+3(1-2w\partial_w)\partial_\nu\right) Q_\nu
     \Bigr]_{\nu=\frac{\sqrt{33}-3}{6}},
\end{split}
\end{align}
where $ c,c',d,d' $ are constants.
At $ w = 1 $, they are regular only if $ d = d' = 0 $.
Then, to remove divergence of at $ w = -1 $, we also obtain $ c = c' = 0 $.
Therefore $ \mu^2 = -1/3 $ mode does not exist.

All the KK modes have positive $ \mu^2 $ hence we conclude the spacetime is stable against linear scalar perturbation.

%%%%%%%%%%%%%%%%%%%%%%%%%%%%%%%%%%%%%%%%%%%%%%%%%%%%%%%%%%%%%%%%%%%%%%%%
\section{
Conclusion and discussion
}
\label{sec:concl}

In this paper, we have formulated an adequate boundary condition for linear perturbations of axisymmetric codimension-2 braneworlds, extending the doubly covariant formulation of \cite{Mukohyama:2000ga} for codimension-1 braneworlds.
An advantage of this formulation is that it allows much more freedom for convenient gauge choices than others.
The boundary condition has been expanded in perturbations and resulted in equation (\ref{eq:bc.perturb}), where $ \delta\mcl_{0,1,2} $ and $ \delta\mca $ are given by (\ref{eq:deltaL0.simple}), (\ref{eq:deltaL1.simple}), (\ref{eq:deltaL2.simple}), and (\ref{eq:deltaA.simple}).

As an application, we have investigated linear perturbations of a six-dimensional braneworld model in section \ref{sec:appl}.
We have derived the boundary condition for scalar-type perturbations that was used in our previous analysis \cite{Yoshiguchi:2005nn}, where the dynamical stability of the model with Minkowski branes was shown numerically.
In the present paper, moreover, we have analytically shown the stability of the system in the limit where the warp factor becomes unity.

The present status of our understanding of gravity in codimension-2 braneworlds is still premature and indicates that more investigations are necessary.
Our future work in this direction has at least two branches.
One is the extension of our formalism to more general setup where the geometry in the bulk is not axisymmetric and/or where general matter contents other than tension are put on the branes.
In particular, finite thickness of the brane will take important roles in this context \cite{deRham:2005ci}.
It is expected that the extension will make it possible to reveal more properties of gravity in the warped flux compactification.
The other is to investigate the nature of the geometry in the high-energy regime.
The importance of the solutions with de Sitter branes comes from its connections to the inflationary early universe and/or the present accelerating expansion.
In \cite{Mukohyama:2005yw}, a sign of instability was noticed when the Hubble expansion rate on the brane is sufficiently large.
In a forthcoming publication we hope to investigate the possibility of such an instability and to understand its origin.
It is also worthwhile to consider a black hole on the brane \cite{Kinoshita:2006} and investigate its stability.

Further developments are needed also in the context of quantum dynamics of the braneworld.
In particular, quantization of matter contents is to give important predictions of inflation in the early universe, and the most direct way of quantization requires the action of the system.
In this respect, it is preferable if we can obtain a doubly covariant action including not only fields on the brane and those in the bulk but also the brane position as an independent variable.
It had actually been believed for a long time that the action principle of singular surfaces (such as branes) would not allow variations of the position of the surface \cite{Friedman:1997fu,Gladush:2000rs}.
If this folklore were true then it would be impossible to find an action principle in a doubly covariant form.
In the codimension-1 case, however, this difficulty was ameliorated and actually resolved by introduction of Lagrange multipliers imposing continuity of induced metric \cite{Mukohyama:2001pb}.
It should be possible and worthwhile to extend the action principle to the codimension-2 case.

%%%%%%%%%%%%%%%%%%%%%%%%%%%%%%%%%%%%%%%%%%%%%%%%%%%%%%%%%%%%%%%%%%%%%%%%
\begin{acknowledgments}
The authors acknowledge helpful discussions with Hiroyuki Yoshiguchi on the early stage of this study.
The authors thank Katsuhiko Sato for his continuing encouragement.
A part of this work was done during SM's visit to University of Victoria.
He is grateful to Werner Israel for his warm hospitality and stimulating discussions.
The authors also thank him for comments on the manuscript.
This work was in part supported by JSPS through a Grant-in-Aid for JSPS Fellows (YS) and by MEXT through a Grant-in-Aid for 21st Century COE Program `Quantum Extreme Systems and Their Symmetries' (SK) and a Grant-in-Aid for Young Scientists (B) No.~17740134 (SM).
\end{acknowledgments}

\appendix
%%%%%%%%%%%%%%%%%%%%%%%%%%%%%%%%%%%%%%%%%%%%%%%%%%%%%%%%%%%%%%%%%%%%%%%%
\section{
Harmonics in Minkowski space
}
\label{app:harmonics}

It is useful to decompose the perturbed quantities by the scalar, vector, and tensor harmonics in Minkowski space.
For the time being, the number of dimensions is taken to be $ n $.
$ \eta_{\mu\nu} \mathrm dx^\mu \mathrm dx^\nu $ denotes $ n $-dimensional Minkowski metric.
Indices are raised or lowered by $ \eta $.

The scalar harmonics are defined as $ Y = \exp(\mathrm i k_\mu x^\mu) $, by which any scalar function of the position in the Minkowski space, $ f(x) $, can be expanded as
\begin{equation}
f = \int\mathrm dk c(k) Y,
\end{equation}
where $ \mathrm dk \equiv \prod_{\mu=0}^{n-1} \mathrm dk_\mu $.

The vector harmonics are defined as
\begin{equation}
V_{(\mathrm T)\mu}
 = u_\mu Y, \quad
V_{(\mathrm L)\mu}
 = \partial_\mu Y = \mathrm i k_\mu Y,
\end{equation}
where an arbitrary constant vector $ u_\mu $ satisfies $ k^\mu u_\mu = 0 $ for non-null $ k_\mu $, i.e., $ k^\mu k_\mu \neq 0 $.
For null $ k_\mu $, it satisfies $ \tau^\mu u_\mu = 0 $ for an arbitrary timelike vector $ \tau^\mu $ in addition to the above condition.
By the harmonics, any vector field $ v^\mu(x) $ can be expanded as
\begin{equation}
v_\mu
 = \int\mathrm dk
   \left(c_{(\mathrm T)}(k) V_{(\mathrm T)\mu}
   + c_{(\mathrm L)}(k) V_{(\mathrm L)\mu}\right),
\end{equation}
where the transverse part satisfies $ \partial^\mu \int\mathrm dk c_{(\mathrm T)} V_{(\mathrm T)\mu} = 0 $.

The tensor harmonics are defined as
\begin{align}
\begin{split}
T_{(\mathrm T)\mu\nu}
 & = s_{\mu\nu} Y, \\
T_{(\mathrm{TL})\mu\nu}
 & = \partial_\mu V_{(\mathrm T)\nu} + \partial_\nu V_{(\mathrm T)\mu}
   = \mathrm i \left(u_\mu k_\nu + u_\nu k_\mu\right) Y, \\
T_{(\mathrm{LL})\mu\nu}
 & = \partial_\mu V_{(\mathrm L)\nu} + \partial_\nu V_{(\mathrm L)\mu}
     - \frac{2}{n} \partial^\sigma V_{(\mathrm L)\sigma} \eta_{\mu\nu}
   = \left(-2 k_\mu k_\nu
     + \frac{2}{n} k^\sigma k_\sigma \eta_{\mu\nu}\right) Y, \\
T_{(\mathrm Y)\mu\nu}
 & = \eta_{\mu\nu} Y.
\end{split}
\end{align}
$ s_{\mu\nu} $ is an arbitrary transverse-traceless constant symmetric tensor that satisfies $ k^\mu s_{\mu\nu} = s^\mu{}_\mu = 0 $ for non-null $ k_\mu $, i.e., $ k^\mu k_\mu \neq 0 $.
For null $ k_\mu $, it satisfies $ \tau^\mu s_{\mu\nu} = 0 $ for a timelike vector $ \tau^\mu $ in addition to the above condition.
By the harmonics, any tensor field $ t_{\mu\nu}(x) $ can be expanded as
\begin{equation}
t_{\mu\nu}
 = \int\mathrm dk
   \left(c_{(\mathrm T)}(k) T_{(\mathrm T)\mu\nu}
         + c_{(\mathrm{TL})}(k) T_{(\mathrm{TL})\mu\nu}
         + c_{(\mathrm{LL})}(k) T_{(\mathrm{LL})\mu\nu}
         + c_{(\mathrm Y)}(k) T_{(\mathrm Y)\mu\nu}\right).
\end{equation}

%%%%%%%%%%%%%%%%%%%%%%%%%%%%%%%%%%%%%%%%%%%%%%%%%%%%%%%%%%%%%%%%%%%%%%%%
\section{
Bulk gauge transformations
}
\label{app:gauge}

For our analysis it is useful to decompose the perturbations according to the transformation property from the viewpoint of four-dimensional space.
Now all the geometrical quantities are decomposed by the harmonics on the Minkowski space summarized in Appendix~\ref{app:harmonics}.
The perturbations of the metric and gauge potential are, respectively, decomposed as 
\begin{align}
\begin{split}
\delta g_{MN} \mathrm dx^M \mathrm dx^N
 & = \int\mathrm dk \bigl(
     \left(h_{(\mathrm T)} T_{(\mathrm T)\mu\nu}
     + h_{(\mathrm{TL})} T_{(\mathrm{TL})\mu\nu}
     + h_{(\mathrm{LL})} T_{(\mathrm{LL})\mu\nu}
     + h_{(\mathrm Y)} T_{(\mathrm Y)\mu\nu}\right) \mathrm dx^\mu \mathrm dx^\nu
     \bigr. \\
 & \qquad
     \bigl.
     + \left(h_{(\mathrm T)a} V_{(\mathrm T)\mu}
     + h_{(\mathrm L)a} V_{(\mathrm L)\mu}\right) \mathrm dx^a \mathrm dx^\mu
     + h_{ab} Y \mathrm dx^a \mathrm dx^b
     \bigr) \\
\delta A_M \mathrm dx^M
 & = \int\mathrm dk \left(
     \left(a_{(\mathrm T)} V_{(\mathrm T)\mu}
           + a_{(\mathrm L)} V_{(\mathrm L)\mu}\right) \mathrm dx^\mu
     + a_a Y \mathrm dx^a
     \right).
\end{split}
\end{align}
Indication of integration is often omitted in the main text.

By the double covariance shown in the main text, we can arbitrarily transform the bulk gauge.
Infinitesimal coordinate transformation and $ U(1) $ gauge transformation for the perturbations are, respectively, denoted as
\begin{equation}
\delta g_{MN}
 \to \delta g_{MN} - \nabla_M\xi_N - \nabla_N\xi_M, \quad
\delta A_M
 \to \delta A_M + \partial_M \chi.
\end{equation}
Gauge parameters are decomposed with the harmonics as follows
\begin{align}
\begin{split}
\xi_M \mathrm dx^M
 & = \int\mathrm dk
     \left(\left(\tilde\xi_{(\mathrm T)} V_{(\mathrm T)\mu}
     + \tilde\xi_{(\mathrm L)} V_{(\mathrm L)\mu}\right) \mathrm dx^\mu
     + \tilde\xi_a Y \mathrm dx^a\right), \\
\chi
 & = \int\mathrm dk \tilde\chi Y.
\end{split}
\end{align}
We assume that the bulk perturbations are axisymmetric.
In such a case $ \tilde\xi $'s can only depend on $ r $.
Note, however, that the $ U(1) $ gauge parameter $ \tilde\chi $ still can vary with $ \phi $.
By these gauge parameters, the metric components are transformed in accordance with their types.

The tensor part is gauge-invariant
\begin{equation}
h_{(\mathrm T)} \to \bar h_{(\mathrm T)} = h_{(\mathrm T)}.
\end{equation}

As for the vector part, components are transformed as
\begin{align}
\begin{split}
h_{(\mathrm{TL})}
 & \to \bar h_{(\mathrm{TL})} = h_{(\mathrm{TL})} - \tilde\xi_{(\mathrm T)}, \\
h_{(\mathrm T)r}
 & \to \bar h_{(\mathrm T)r}
   = h_{(\mathrm T)r}
     - \tilde\xi_{(\mathrm T)}' + \frac{2}{r} \tilde\xi_{(\mathrm T)}, \\
h_{(\mathrm T)\phi}
 & \to \bar h_{(\mathrm T)\phi} = h_{(\mathrm T)\phi}, \\
a_{(\mathrm T)}
 & \to \bar a_{(\mathrm T)} = a_{(\mathrm T)},
\end{split}
\end{align}
where the prime denotes derivative with respect to $ r $.
Setting $ \tilde\xi_{(\mathrm T)} = h_{(\mathrm{TL})} $ gives $ \bar h_{(\mathrm{TL})} = 0 $, which is the gauge condition utilized in section \ref{sec:appl.stab.vector}.

As for the scalar part, components are transformed as
\begin{align}
\begin{split}
h_{(\mathrm{LL})}
 & \to \bar h_{(\mathrm{LL})} = h_{(\mathrm{LL})} - \tilde\xi_{(\mathrm L)}, \\
h_{(\mathrm Y)}
 & \to \bar h_{(\mathrm Y)}
   = h_{(\mathrm Y)} + \frac{1}{2} k^\mu k_\mu \tilde\xi_{(\mathrm L)}
     - 2 r f \tilde\xi_r, \\
h_{(\mathrm L)r}
 & \to \bar h_{(\mathrm L)r}
   = h_{(\mathrm L)r}
     - \tilde\xi_{(\mathrm L)}' + \frac{2}{r} \tilde\xi_{(\mathrm L)}
     - \tilde\xi_r, \\
h_{(\mathrm L)\phi}
 & \to \bar h_{(\mathrm L)\phi} = h_{(\mathrm L)\phi} - \tilde\xi_\phi, \\
h_{rr}
 & \to \bar h_{rr} = h_{rr} - 2 \tilde\xi_r' - \frac{f'}{f} \tilde\xi_r, \\
h_{r\phi}
 & \to \bar h_{r\phi} = h_{r\phi} - \tilde\xi_\phi'
                       + \frac{f'}{f} \tilde\xi_\phi, \\
h_{\phi\phi}
 & \to \bar h_{\phi\phi} = h_{\phi\phi} - f f' \tilde\xi_r, \\
a_{(\mathrm L)}
 & \to \bar a_{(\mathrm L)} = a_{(\mathrm L)}
                            + \tilde\chi - A \frac{\tilde\xi_\phi}{f}, \\
a_r
 & \to \bar a_r = a_r + \partial_r \tilde\chi
                  - A \left(\frac{\tilde\xi_\phi}{f}\right)', \\
a_\phi
 & \to \bar a_\phi = a_\phi + \partial_\phi \tilde\chi - A' f \tilde\xi_r.
\end{split}
\end{align}
Setting
\begin{align}
\begin{split}
\tilde\xi_{(\mathrm L)}
 & = h_{(\mathrm{LL})}, \\
\tilde\xi_r
 & = h_{(\mathrm L)r} + \frac{2}{r} h_{(\mathrm{LL})} - h_{(\mathrm{LL})}', \\
\tilde\xi_\phi
 & = f \int_{C'}^r \mathrm dr' \frac{h_{r\phi}(r')}{f(r')}, \\
\tilde\chi
 & = - a_{(\mathrm L)} + A \frac{\tilde\xi_\phi}{f},
\end{split}
\end{align}
with $ C' $ an arbitrary constant gives $ \bar h_{(\mathrm{LL})} = \bar h_{(\mathrm L)r} = \bar h_{r\phi} = \bar a_{(\mathrm L)} = 0 $.
This is the gauge condition used in section \ref{sec:appl.stab.scalar} together with the following redefinitions
\begin{equation}
\bar h_{(\mathrm Y)} \equiv \Psi, \quad
\bar h_{(\mathrm L)\phi} \equiv h_{(\mathrm L)\phi}, \quad
\bar h_{rr} \equiv (\Phi_1 + \Phi_2)/f, \quad
\bar h_{\phi\phi} \equiv - (\Phi_1 + 3 \Phi_2) f, \quad
\bar a_r \equiv a_r, \quad
\bar a_\phi \equiv a_\phi.
\end{equation}

Note that the gauge transformation considered in this appendix is for modes with $ k^{\mu} \neq 0 $.
Modes with $ k^{\mu} = 0 $ can be included in the background without loss of generality.

%%%%%%%%%%%%%%%%%%%%%%%%%%%%%%%%%%%%%%%%%%%%%%%%%%%%%%%%%%%%%%%%%%%%%%%%
\section{
The $ \alpha \to 1 $ limit
}
\label{app:sphere}

When the four-dimensional geometry is the Minkowski, the spacetime is
given by
\begin{equation}
g_{MN} \mathrm dx^M \mathrm dx^N
 = r^2 \eta_{\mu\nu} \mathrm dx^\mu \mathrm dx^\nu
   + \frac{\mathrm dr^2}{f(r)} + f(r) \mathrm d\phi^2, \quad
A_M \mathrm dx^M
 = A(r) \mathrm d\phi,
\end{equation}
where
\begin{equation}
f
 = - \frac{\Lambda_6}{10} r^2 - \frac{\mu_b}{r^3} - \frac{b^2}{12 r^6}, \quad
A
 = \frac{b}{3 r^3}.
\end{equation}
Parameters $ (\mu_b,b) $ are replaced by the roots $ (r_+,r_-) $ satisfying $ f(r_\pm) = 0 $.
In the $ r_-/r_+ \to 1 $ limit, the metric function becomes
\begin{equation}
f
 = 2 \Lambda_6 \epsilon^2 \left(1 - \left(\frac{r-r_0}{\epsilon}\right)^2\right)
   + \mco(r-r_0)^3,
\end{equation}
where $ r_0 \equiv (r_++r_-)/2 $ and $ \epsilon \equiv (r_+-r_-) / 2 $.
Then a coordinate transformation 
\begin{equation}
w = \frac{r-r_0}{\epsilon}, \quad
\varphi = 2 \Lambda_6 \epsilon \phi
\end{equation}
gives the following representation to the metric:
\begin{equation}
g_{MN} \mathrm dx^M \mathrm dx^N
 \simeq r_0^2 \eta_{\mu\nu} \mathrm dx^\mu \mathrm dx^\nu
   + \frac{1}{2\Lambda_6}
     \left(\bar f(w) \mathrm dw^2 + \frac{\mathrm d\varphi^2}{\bar f(w)}\right),
\end{equation}
where $ \bar f(w) \equiv 1-w^2 $.
The geometry of the extra space is locally $ 2 $-sphere with the radius $ R \equiv 1/\sqrt{2\Lambda_6} $.
The gauge potential is transformed as
\begin{equation}
A \mathrm d\phi
 = A \frac{\mathrm d\phi}{\mathrm d\varphi} \mathrm d\varphi
 = \frac{b}{3 (r_0+\epsilon w)^3} \frac{R^2}{\epsilon} \mathrm d\varphi
 = \left(\frac{R r_0}{3 \epsilon} - R w
         + \mco(\epsilon)\right) \mathrm d\varphi.
\end{equation}
The constant can be always removed by some $ U(1) $ gauge transformation.
Then, in the $ \epsilon \to 0 $ limit, we obtain
\begin{equation}
A_M \mathrm dx^M
 \simeq R \bar A(w) \mathrm d\varphi,
\end{equation}
where $ \bar A(w) \equiv -w $.

%%%%%%%%%%%%%%%%%%%%%%%%%%%%%%%%%%%%%%%%%%%%%%%%%%%%%%%%%%%%%%%%%%%%%%%%
\section{
Regular quantities
}
\label{app:regular}

In this Appendix, we show independent regular quantities of each type of perturbations on the $ \alpha = 1 $ background.
See Appendix~\ref{app:gauge} for our gauge choices.
These quantities are evaluated in the inertial frame on the brane and the values are required to be regular.
In our position, regularity has been considered to work complementarily as boundary conditions.

As for tensor-type perturbation, independent quantities in the inertial
frame on the boundary are $ h_{(\mathrm T)} $ itself and, for example, 
\begin{equation}
(e_\mu)^M (e_w)^P (e_\nu)^N (e_w)^Q \delta C_{MPNQ}
 = -\frac{1}{2}
    \left(w h_{(\mathrm T)}' + \left(\frac{3}{10}-m^2\right)
          h_{(\mathrm T)}\right) T_{(\mathrm T)\mu\nu},
\end{equation}
where $ \{e_a\} $ denotes (unperturbed) vielbein.
From this combination, we deduce that
\begin{equation}
h_{(\mathrm T)}, \quad h_{(\mathrm T)}'
\end{equation}
must be regular at the boundaries.

As for vector-type perturbations, independent quantities in the inertial frame on the boundaries are, after using equations of motion,
\begin{align}
\begin{split}
(e_\mu)^M (e_w)^P (e_\nu)^N (e_\varphi)^Q \delta C_{MPNQ}
 & = \frac{w}{1-w^2} h_{(\mathrm T)\varphi} T_{(\mathrm{TL})\mu\nu}
     + \frac{1}{2} h_{(\mathrm T)\varphi}' \partial_\nu V_{(\mathrm T)\mu}, \\
(e_\mu)^M (e_\nu)^N \delta F_{MN}
 & = a_{(\mathrm T)}
     (\partial_\mu V_{(\mathrm T)\nu} - \partial_\nu V_{(\mathrm T)\mu}), \\
(e_w)^M (e_\mu)^N \delta F_{MN}
 & = \sqrt{1-w^2} a_{(\mathrm T)}' V_{(\mathrm T)\mu}.
\end{split}
\end{align}
From this combination, we deduce that
\begin{equation}
\frac{h_{(\mathrm T)\varphi}}{1-w^2}, \quad
h_{(\mathrm T)\varphi}', \quad
a_{(\mathrm T)}, \quad
\sqrt{1-w^2} a_{(\mathrm T)}'
\end{equation}
must be regular at the boundaries.

As for scalar-type perturbations, independent quantities in the inertial frame on the boundaries are, after using equations of motion,
\begin{align}
\begin{split}
\delta R
 & = -\frac{1}{2} (\mu^2 \Phi_1 + 4 w \Phi_2') Y, \\
\delta\left(R^{MN}{}_{M'N'} R^{M'N'}{}_{MN}\right)
 & = -2 [\mu^2 (5\Phi_1+6\Phi_2) + 20 w \Phi_2'] Y, \\
(e_w)^M (e_\mu)^P (e_w)^N (e_\nu)^Q \delta C_{MPNQ}
 & = -\frac{1}{2} (\Phi_1+\Phi_2) Y_{,\mu\nu} \quad (\mu \neq \nu), \\
(e_\mu)^M (e_\rho)^P (e_\nu)^N (e_w)^Q \delta C_{MPNQ}
 & = -\frac{1}{8} \left(\sqrt{1-w^2} (\Phi_1'+4\Phi_2')
     - \frac{2w}{\sqrt{1-w^2}} (\Phi_1+2\Phi_2)\right)
       \eta_{\mu\nu} Y_{,\rho} \quad (\mu,\nu \neq \rho).
\end{split}
\end{align}
From this combination, we deduce that
\begin{equation}
\Phi_2, \quad
\Phi_2', \quad
\Phi_1, \quad
\sqrt{1-w^2} \Phi_1' - \frac{2w}{\sqrt{1-w^2}} (\Phi_1+2\Phi_2)
\end{equation}
must be regular at the boundaries.

%%%%%%%%%%%%%%%%%%%%%%%%%%%%%%%%%%%%%%%%%%%%%%%%%%%%%%%%%%%%%%%%%%%%%%%%
\section{
Action for the background boundary condition
}
\label{app:action}

In this Appendix we derive the action for a codimension-2 brane.
We start with a general axisymmetric metric 
\begin{equation}
g_{MN} \mathrm dx^M \mathrm dx^N
 = \hat g_{\alpha\beta}(\hat x)
   \mathrm d\hat x^\alpha \mathrm d\hat x^\beta
   + L(\hat x)^2 \mathrm d\phi^2,
\end{equation}
where, throughout this Appendix, $ \alpha $ and $ \beta $ take $ \alpha,\beta = 0,\ldots,D $ and $ \hat x^\alpha $ does not include $ \phi $.
The axis of rotation is defined by $ L(\hat x) = 0 $ and the period of the angular coordinate $ \phi $ is $ \Delta\phi $.
We define $ n_M $ as the out-directed unit normal to each constant $ L $ surface
\begin{equation}
n_M
 = \frac{\nabla_M L}{\sqrt{\nabla_M L \nabla^M L}}
 = \frac{\partial_M L}{\partial_\perp L},
\end{equation}
where $ \partial_\perp L = n^M \partial_M L $.
Introducing radial coordinate by $ \mathrm d\rho = n_M \mathrm dx^M $, the $ (D+1) $-dimensional metric $ \hat g $ further admits a decomposition
\begin{equation}
\hat g_{\alpha\beta} \mathrm d\hat x^\alpha \mathrm d\hat x^\beta
 = q_{\mu\nu}(y,\rho) \mathrm dy^\mu \mathrm dy^\nu + \mathrm d\rho^2.
\end{equation}
It is clear that the above defined vector field $ n_M $ is identical with that given in the main part as orthogonality to other vectors is reproduced.
Further, since $ \varphi^M n_M \propto \partial L/\partial \phi = 0 $ and $ {e_\mu}^M n_M \propto \partial L/\partial y^\mu = 0 $, we see that $ L $ only depends on $ \rho $.
This means that each hypersurface of constant $ L $ corresponds to that of constant $ \rho $.
We shift the origin of $ \rho $ without loss of generality so that $ \rho \to 0+ $ corresponds to $ L(\rho) \to 0+ $.

Now a neighborhood of the axis of the symmetry $ L = 0 $ is foliated by an one parameter family of $ (D+1) $-dimensional hypersurfaces $ \Sigma_\ell $ each placed on constant $ \rho = \ell > 0 $.
We also define $ D $-dimensional submanifolds $ \mcb_\ell $ contained in $ \Sigma_\ell $ by restricting $ \phi = \text{const} $.
The metric of $ \mcb_\ell $ is $ q_{\mu\nu}(y,\ell) $.

We would like to place a codimension-2 brane at $ \rho = 0 $ but, in order to regularize the system, first consider the hypersurface $ \Sigma_\ell $ and define the physical brane as the $ \ell \to 0 $ limit of $ \Sigma_\ell $.
In this limit the induced metric on the brane is given by $ q_{\mu\nu} \equiv q_{\mu\nu}(y,0) $.
Thus, we assume that $ q_{\mu\nu} $ and its curvature tensors remain regular at $ \rho = 0 $.

Let us consider one such hypersurface $ \Sigma_\ell $.
The whole manifold $ \mcm $ is seen as union of two parts $ \mcm = \mcm^+ \cup \mcm^- $ such that $ \mcm^+ $ and $ \mcm^- $ corresponding to $ \rho > \ell $ and $ \rho < \ell $, respectively.
The Einstein--Hilbert action is decomposed to have two Gibbons--Hawking terms associated with two sides of $ \Sigma_\ell $ as
\begin{align}
S_\mathrm{EH}
 & = \frac{M_{D+2}^D}{2} \int_\mcm \mathrm d^{D+2}x \sqrt{-g} R \nonumber\\
 & = \frac{M_{D+2}^D}{2} \left(
     \int_{\mcm^+}
     \mathrm d^{D+2}x \sqrt{-g} R
     - 2 \int_{\partial\mcm^+}
         \mathrm d^{D+1}\hat x \sqrt{-q} L K
     + \int_{\mcm^-}
       \mathrm d^{D+2}x \sqrt{-g} R
     + 2 \int_{\partial\mcm^-}
         \mathrm d^{D+1}\hat x \sqrt{-q} L K
     \right).
\end{align}
Note that in this case the direction of the normal is from $ \mcm^- $ to $ \mcm^+ $.
Trace of the extrinsic curvature of a constant $ \rho $ surface is given by
\begin{equation}
K
 = \frac{1}{\sqrt{-g}} \partial_M \left(\sqrt{-g} n^M\right)
 = \frac{\partial_\perp L}{L}
   + \frac{\partial_\perp \sqrt{-q}}{\sqrt{-q}}
 \equiv {}^{(1)}K + \hat K.
\end{equation}
In the $ \ell \to 0+ $ limit, the Gibbons--Hawking term of $ \partial\mcm^+ $ has the following limit
\begin{equation}
\int_{\partial\mcm^+} \mathrm d^{D+1}\hat x \sqrt{-q} L K
 = \int_{\mcb_\ell} \mathrm d^Dy \sqrt{-q}
   \int_0^{\Delta\phi} \mathrm d\phi L \left({}^{(1)}K + \hat K\right)
 \to \int_\mcb \mathrm d^Dy \sqrt{-q} \Delta\phi \partial_\perp L,
\end{equation}
where we regarded $ \hat K $ as a finite quantity in the limit.
On the other hand, the topological terms associated with $ \mcm^- $ and $ \partial\mcm^- $ give
\begin{multline}
\int_{\mcm^-} \mathrm d^{D+2}x \sqrt{-g} R
+ 2 \int_{\partial\mcm^-} \mathrm d^{D+1}\hat x \sqrt{-q} L K \\
 = \int_{\mcb_\ell} \mathrm d^Dy \sqrt{-q}
   \left(\int_0^{\Delta\phi} \mathrm d\phi \int_\ell \mathrm d\rho L
         \left({}^{(2)}R + \hat R\right)
         + 2 \int_0^{\Delta\phi} \mathrm d\phi L
           \left({}^{(1)}K + \hat K\right)\right),
\end{multline}
where we have decomposed the $ (D+2) $-dimensional Ricci scalar $ R $ into two-dimensional $ (\rho,\phi) $ part $ {}^{(2)}R $ and other parts $ \hat R $.
In the $ \ell \to 0+ $ limit $ {}^{(2)}R \sim \mco(1/L) $, whereas $ \hat R \sim \mco(1) $ because of the regularity of the induced metric $ q_{\mu\nu}(y,\rho=0) $ on the brane.

Therefore, by the Gauss--Bonnet theorem, terms in the square brackets give
\begin{equation}
\int_0^{\Delta\phi} \mathrm d\phi \int_\ell \mathrm d\rho L
\left({}^{(2)}R + \hat R\right)
+ 2 \int_0^{\Delta\phi} \mathrm d\phi L \left({}^{(1)}K + \hat K\right)
 \to \iint \mathrm d\phi \mathrm d\rho L {}^{(2)}R
     + 2 \int \mathrm d\phi L {}^{(1)}K
 = 4 \pi.
\end{equation}
Then the Einstein--Hilbert action becomes
\begin{equation}
S_\mathrm{EH}
 = \frac{M_{D+2}^D}{2} \left(
   \int_\mcm \mathrm d^{D+2}x \sqrt{-g} R
 + \int_\mcb \mathrm d^Dy \sqrt{-q}
   \left(-2 \Delta\phi \partial_\perp L + 4 \pi\right)
   \right).
\end{equation}
The matter action is of the tension of the brane
\begin{equation}
S_\mathrm{matter}
 = - \int_\mcb \mathrm d^Dy \sqrt{-q} \sigma.
\end{equation}
Variation $ q^{\mu\nu} \to q^{\mu\nu} + \delta q^{\mu\nu} $ to the total action $ S = S_\mathrm{EH} + S_\mathrm{matter} $ gives the boundary condition as an equation of motion
\begin{equation}
- 2 \Delta\phi \partial_\perp L + 4 \pi - \frac{2 \sigma}{M_{D+2}^D} = 0,
\end{equation}
which is actually the same as (\ref{eq:dL}).

%%%%%%%%%%%%%%%%%%%%%%%%%%%%%%%%%%%%%%%%%%%%%%%%%%%%%%%%%%%%%%%%%%%%%%%%
\bibliographystyle{hunsrt}
\bibliography{boundary}

\end{document}